\documentclass[12pt]{article}
\usepackage{authblk} 
\usepackage{algorithm}
\usepackage{algorithmic}
\usepackage{afterpage}
\usepackage{booktabs}
\usepackage{lipsum}
\usepackage{float}
\usepackage{float,lscape}
\usepackage[utf8]{inputenc}
\usepackage[T1]{fontenc}
\usepackage{amsmath,amsfonts,amssymb,amsthm}
\usepackage{stmaryrd}
\usepackage{lscape}
\usepackage{hyperref}
\usepackage[a4paper, total={6in, 8in}]{geometry}
\usepackage{multirow}
\usepackage{multirow, tabularx}
 \usepackage[version=4]{mhchem}
\usepackage{graphicx}
\usepackage[export]{adjustbox}
\graphicspath{ {./images/} }
\newtheorem{theorem}{Theorem}

\newtheorem{example}[theorem]{Example}

\newcommand{\nN}{\scriptscriptstyle  N}

\title{ Neutrosophic Birnbaum-Saunders distribution with applications
}
\author[1]{Mansooreh Razmkhah 
}
\author[1,2]{Mohammad Arashi}
\author[3]{Andriette Bekker}
\author[4]{Filipe J. Marques}
\affil[1]{\footnotesize{Department of Statistics, Faculty of Mathematical Sciences, Ferdowsi University of Mashhad, Iran}}
\affil[2]{\footnotesize{Department of Statistics, Uinversity of Pretoria, Pretoria, South Africa}}
\affil[3]{\footnotesize{Centre for Environmental Studies, Department of Geography, Geoinformatics and Meteorology, University of Pretoria}}
\affil[4]{\footnotesize{Department of Mathematics and Center for Mathematics and Applications (NOVA Math), NOVA School of Science and Technology, Portugal}}
\date{}

\begin{document}
\maketitle

\begin{abstract}

Classical statistics deals with determined and precise data analysis. But in reality, there are many cases where the information is not accurate and a degree of impreciseness, uncertainty, incompleteness, and vagueness is observed. 
In these situations, uncertainties can make classical statistics less accurate. That is where neutrosophic statistics steps in to improve accuracy in data analysis.
In this article, we consider the Birnbaum-Saunders distribution (BSD) which is very flexible and practical for real world data modeling. 
By integrating the neutrosophic concept, we improve the BSD's ability to manage uncertainty effectively. 
In addition, we provide maximum likelihood parameter estimates. Subsequently, we illustrate the practical advantages of the neutrosophic model using two cases from the industrial and environmental fields. This paper emphasizes the significance of the neutrosophic BSD as a robust solution for modeling and analysing imprecise data, filling a crucial gap left by classical statistical methods.

\end{abstract}

KEYWORDS: Birnbaum-Saunders distribution, environmental neutrosophic random variables, maximum likelihood estimation, neutrosophic distribution.

\section{Introduction}

\subsection{Contextualizing the problem}
Neutrosophic theory provides a framework for dealing with uncertainty in situations where the data or a part of it are indeterminate to some degrees. 
In traditional mathematics, certainty is the most crucial necessity, however  real data show a greater amount of fuzziness, uncertainty, and incompleteness compared to the presence of determinate information.  Smarandache \cite{Smarandache:2014} introduced the concept of neutrosophic statistics as an evolution of classical statistics. The purpose of this expansion is to manage data that is uncertain or imprecise, together with its associated statistical probability distributions, by using a set of values as an approximation of the original precise values. The practical utility of neutrosophic statistical methods becomes evident in addressing such situations.
It enables us to interpret and organize the neutrosophic data in order to reveal underlying patterns in a more nuanced way than classical methods.

Uncertainty modeling attracts many scientists and engineers because it helps them define and explain the useful information that is hidden in uncertain data. Neutrosophic logic, set, and probability was introduced by Smarandache \cite{Smarandache:1998}. It is a generalization of fuzzy, intuitionistic, Boolean, paraconsistent logics, etc. 
 The neutrosophic concept have been employed to address these inherent uncertainties in a variety of fields such as communication, management and information technology and numerous scientific and engineering disciplines have implemented this logic. Smarandache continued his research, providing further insights into this field, highlighted by Smarandache \cite{Smarandache:2005,Smarandache:2007,Smarandache:2010,Smarandache:2016}, and Smarandache and Pramanik \cite{Smarandache:Pramanik:2016}, and Patro and Smarandache \cite{Patro:Smarandache:2016}.
Then, researchers proceeded to develop this topic. Ali et al. \cite{Ali:2018} investigated the properties of a new concept known as interval complex neutrosophic set. Salama and Alblowi \cite{Salama:Alblowi:2012} introduced definitions of generalized neutrosophic sets and extended the concepts of neutrosophic topological space.

Several studies have focused in recent years on the theory of neutrosophic statistics both methodologically and applied form including correlation, regression analysis, test procedures, time-series, control charts and sampling plans, engineering problems, neutrosophic probability distributions to calculate indeterminacy in real-life problems that produced better results in comparison to classical statistics.  It shifts back to classical statistics when faced with known or certain data. 

Neutrosophic probability distributions have been explored by a few researchers in recent years  and we highlight a few contributions. 
Alhabib et al. \cite{Alhabib:2018} studied the neutrosophic Poisson, exponential, and uniform distributions. 
 A neutrosophic version of the Weibull distribution was developed by Alhasan and Smarandache  \cite{Alhasan:Smarandache:2019}. 
 Khan et al. \cite{Khan:2021} attempted to develop a statistical model of neutrosophic  gamma distribution with applications to complex data analysis. 
Log-normal distribution in neutrosophy concept and its applications in environmental situations was explored by  Khan et al. \cite{khan2021statistical}.
Khan et al. \cite{Khan:2021} introduced the neutrosophic Rayleigh distribution while the neutrosophic generalized Rayleigh distribution was proposed by Norouzirad et al. \cite{Norouzirad:2023}. 
Duan et al. \cite{Duan:2021} examined the applications of  neutrosophic exponential distribution for complex data analysis, and then Rao et al. \cite{rao2023neutrosophic} explored the generalized exponential distribution.
The neutrosophic Beta distribution, along with its properties and applications, was introduced by Sherwani et al. \cite{Sherwani:2021}. 
Shah et al. \cite{Shah:2022} discussed the neutrosophic extension of the Maxwell model. 
 The neutrosophic Kumaraswamy distribution with engineering application was investigated by Ahsan-ul Haq \cite{Ahsan-ul-Haq:2022}. 
Eassa et al. \cite{Eassa:2023} presented the neutrosophic generalized Pareto distribution.

\subsection{Birnbaum-Saunders distribution}

 Out of all statistical distributions, the normal distribution stands out as the most commonly utilized in practical applications. Various new distributions have emerged by modifying the normal distribution through transformations. The two-parameter Birnbaum-Saunders  distribution (BSD) is an example, created by applying a monotone transformation to the standard normal random variable. 
Birnbaum and Saunders  \cite{birnbaum1969new} developed this distribution specifically to model the fatigue life of metals subjected to cyclic stress. Therefore, it is sometimes referred to as the fatigue-life distribution.
In this article,    we focus on a model involves starting a process to adjust the BSD in  the neutrosophic concept, making it more flexible and useful. This approach enables the model to effectively address real-world practical challenges associated with managing uncertain or indeterminate data. 
The ability of the model is to  address situations where data is reported with interval values, offering a robust structure for analysis and decision-making in such complex situations.

 A random variable $T$ follows the Birnbaum-Saunders distribution 
$BS(\alpha, \beta)$ if it can be expressed as
\begin{equation}\label{eq:1}
    T=\beta\left(\frac{\alpha}{2} Z+\sqrt{\left(\frac{\alpha}{2}Z\right)^2+1}\right)^2; ~~~ Z=\frac{1}{\alpha}\left[\sqrt{\frac{T}{\beta}}-\sqrt{\frac{\beta}{T}}\right] \sim \mathcal{N}(0,1),
 \end{equation}
 where $\alpha>0$, and $\beta > 0$ are shape and scale parameters, respectively. The cumulative distribution function (CDF) is defined as
 \begin{equation}\label{eq:cdf}
     F(t;\alpha,\beta) = \Phi\left(\frac{1}{\alpha}\left( \sqrt{\frac{t}{\beta}} -\sqrt{\frac{\beta}{t}}\right)\right), \quad t>0, \quad\alpha,\beta >0 
 \end{equation}
 where $\Phi(\cdot)$  denotes the standard normal distribution CDF. The associated probability density function (PDF) is given by
 \begin{align}\label{eq:pdf}
        f(t;\alpha,\beta)=\frac{1}{\sqrt{2 \pi}} \exp \left\{-\frac{1}{2 \alpha^2}\left( \frac{t}{\beta}+\frac{\beta} {t}-2\right) \right\} \frac{t^{\frac{-3}{2}}(t+\beta)}{2 \alpha \sqrt{\beta}},\nonumber\\ \quad 0<t<\infty, \quad \alpha, \beta >0.
 \end{align}
 
Balakrishnan and Kundu \cite{balakrishnan2019birnbaum} provided a detailed review of the progress of all Birnbaum-Saunders models.

\subsection{Outline of the paper}
 The rest of this article is organized as follows
Section \ref{sec:2} provides a description of neutrosophic Birnbaum-Saunders distribution (NBSD). 
 Section \ref{sec:3} provides an overview of the various statistical characteristics of NBSD. Section \ref{sec:4} elucidates the procedure for estimating neutrosophic parameters. Section \ref{sec:5} performs a simulation study. 
Section \ref{sec:6} 
showcases practical applications of the NBSD with real-world datasets, followed by a comparative analysis with two other established candidate models.
Section \ref{sec:7} 
summarizes the key findings, implications, and practical applications of our study, highlighting the extension of the BSD for data analysis in scenarios of varying indeterminacy and its application in real-world contexts.

\section{Neutrosophic Birnbaum-Saunders distribution}\label{sec:2}
Let $T_{{\scriptscriptstyle N}}\in\left[ T_{\scriptscriptstyle L},T_{\scriptscriptstyle U} \right]$ be a neutrosophic random variable. We assume $T_{\scriptscriptstyle N} \sim NBS(\alpha_{\scriptscriptstyle N}, \beta_{\scriptscriptstyle N})$ and in general say $T_{\scriptscriptstyle N}$ follows a NBSD with neutrosophic shape parameter $\alpha_{{\scriptscriptstyle N}}\in\left[\alpha_{\scriptscriptstyle L}, \alpha_{\scriptscriptstyle U}\right]$ and neutrosophic scale parameter $\beta_{{\scriptscriptstyle N}}\in\left[ \beta_{\scriptscriptstyle L},\beta_{\scriptscriptstyle U} \right]$. Then, the associated neutrosophic PDF and CDF can be expressed as follows respectively 
 
 \begin{align}
 f(t_{\scriptscriptstyle N};\alpha_{\scriptscriptstyle N},\beta_{\scriptscriptstyle N})=\frac{1}{\sqrt{2 \pi}} \exp \left(-\frac{1}{2 {\alpha_{\scriptscriptstyle N}}^2}\left[\frac{t}{\beta_{\scriptscriptstyle N}}+\frac{\beta_{\scriptscriptstyle N}} {t}-2\right]\right) \frac{t_{\scriptscriptstyle N}^{\frac{-3}{2}}[t_{\scriptscriptstyle N}+\beta_{\scriptscriptstyle N}]}{2 \alpha_{\scriptscriptstyle N} \sqrt{\beta_{\scriptscriptstyle N}}}, \nonumber\\ \quad  0<t_{\scriptscriptstyle N}<\infty,  \quad \alpha_{\scriptscriptstyle N}, \beta_{\scriptscriptstyle N} >0 
\end{align}
 and
 \begin{equation}
        F(t_{\scriptscriptstyle N};\alpha_{\scriptscriptstyle N},\beta_{\scriptscriptstyle N})=\Phi \left[\frac{1}{\alpha_{\scriptscriptstyle N}}\left\{\left(\frac{t_{\scriptscriptstyle N}}{\beta_{\scriptscriptstyle N}} \right)^{\frac{1}{2}}-\left(\frac{\beta_{\scriptscriptstyle N}}{t_{\scriptscriptstyle N}}\right)^{\frac{1}{2}}\right\}\right],  \quad  t_{\scriptscriptstyle N}>0, ~~\alpha_{\scriptscriptstyle N}>0,  \quad  \beta_{\scriptscriptstyle N} >0.
 \end{equation}
Within real-world scenarios, every piece of information possesses the capacity to include a certain level of uncertainty, and the values of parameters may likewise be subject to uncertainty.
Obviously, when there is no uncertainty about any of the sample values or parameters, the lower and upper bounds become equal and the model reduces to classical model.

If $T_{\scriptscriptstyle N} \sim \operatorname{NBS}(\alpha_{\scriptscriptstyle N}, \beta_{\scriptscriptstyle N})$, the survival and hazard functions of the NBSD are respectively given by
 \begin{equation}\label{eq:5}
 s(t_{\scriptscriptstyle N};\alpha_{\scriptscriptstyle N},\beta_{\scriptscriptstyle N})= 1-\Phi \left[\frac{1}{\alpha_{\scriptscriptstyle N}}\left\{\left(\frac{t_{\scriptscriptstyle N}}{\beta_{\scriptscriptstyle N}} \right)^{\frac{1}{2}}-\left(\frac{\beta_{\scriptscriptstyle N}}{t_{\scriptscriptstyle N}}\right)^{\frac{1}{2}}\right\}\right]	
 \end{equation}
 and 
 \begin{equation}
 h(t_{\scriptscriptstyle N} ; \alpha_{\scriptscriptstyle N}, \beta_{\scriptscriptstyle N})=\frac{f(t_{\scriptscriptstyle N} ; \alpha_{\scriptscriptstyle N}, \beta_{\scriptscriptstyle N})}{1-F(t_{\scriptscriptstyle N} ; \alpha_{\scriptscriptstyle N}, \beta_{\scriptscriptstyle N})}, \quad t>0 .
 \end{equation}

In order to elucidate the topic more thoroughly, we will provide an illustrative example.

 \begin{example}
 The fatigue lifespan in hours of a ball bearing, $T_{\scriptscriptstyle N}$, follows a NBSD with indeterminate parameters as $T_{\scriptscriptstyle N} \sim NBS([0.08, 0.09],[179.5, 181])$.
 The probability that the fatigue lifespan is more than $170$ hours is calculated from \eqref{eq:5}
\begin{align*}
s(170;\alpha_{\scriptscriptstyle N},\beta_{\scriptscriptstyle N})&=
 1 - \Phi \left[\frac{1}{[0.08, 0.09]}\left\{\left(\frac{170}{[179.5, 181]} \right)^{\frac{1}{2}}-\left(\frac{[179.5, 181]}{170}\right)^{\frac{1}{2}}\right\}\right]\\&=[0.7271649, 0.7834391].
\end{align*}
The minimum and maximum chances for fatigue lifespan to pass $170$ hours are almost $72.7 \%$ and $78.3 \%$ regarding the indeterminacy in the parameters.
 \end{example}
 
\begin{figure}[!htbp]
	\begin{center}
		\begin{tabular}{cc}
			\includegraphics[angle=-90,scale = 0.2]{ 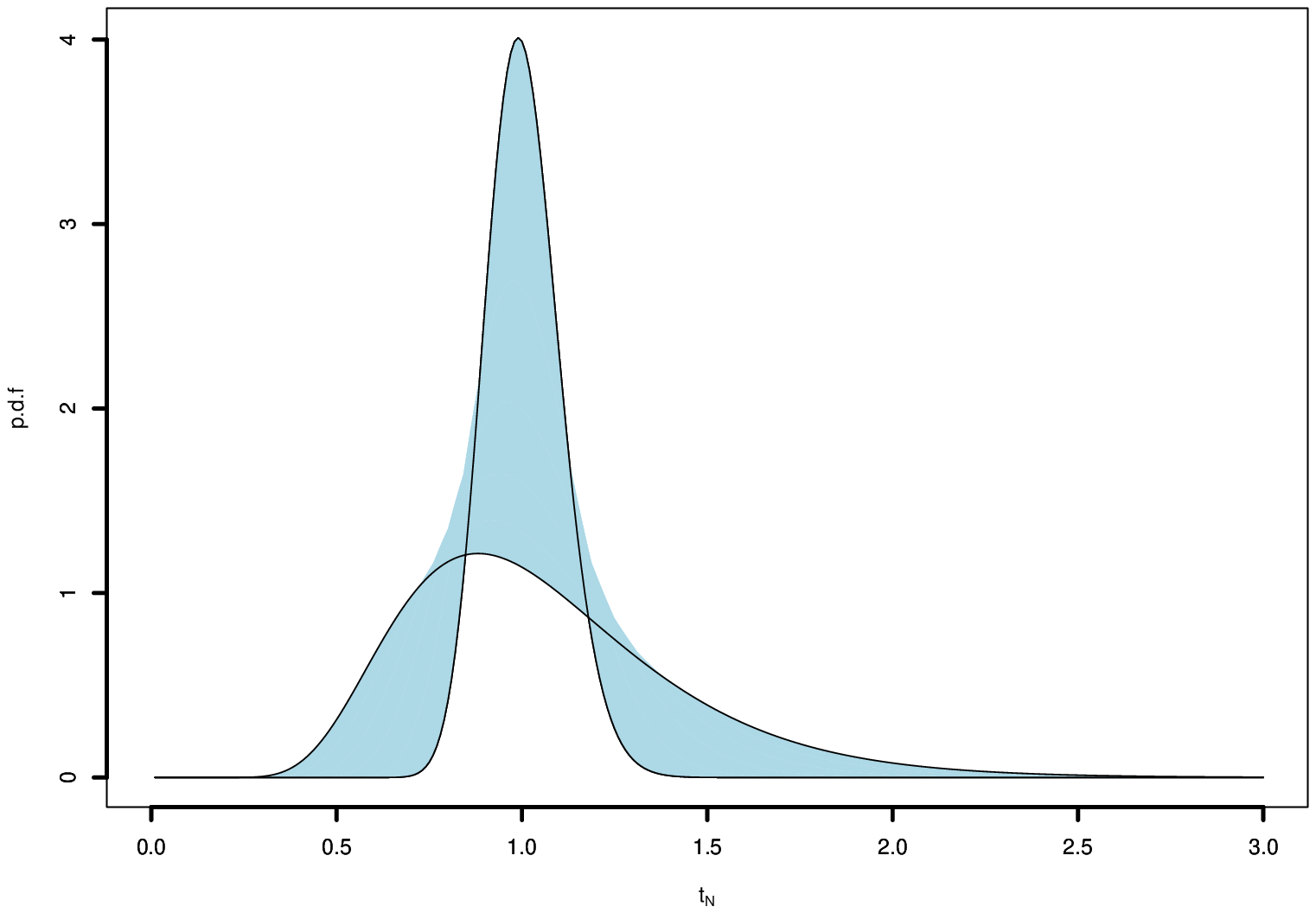}&  
			\includegraphics[angle=-90,scale = 0.2]{ 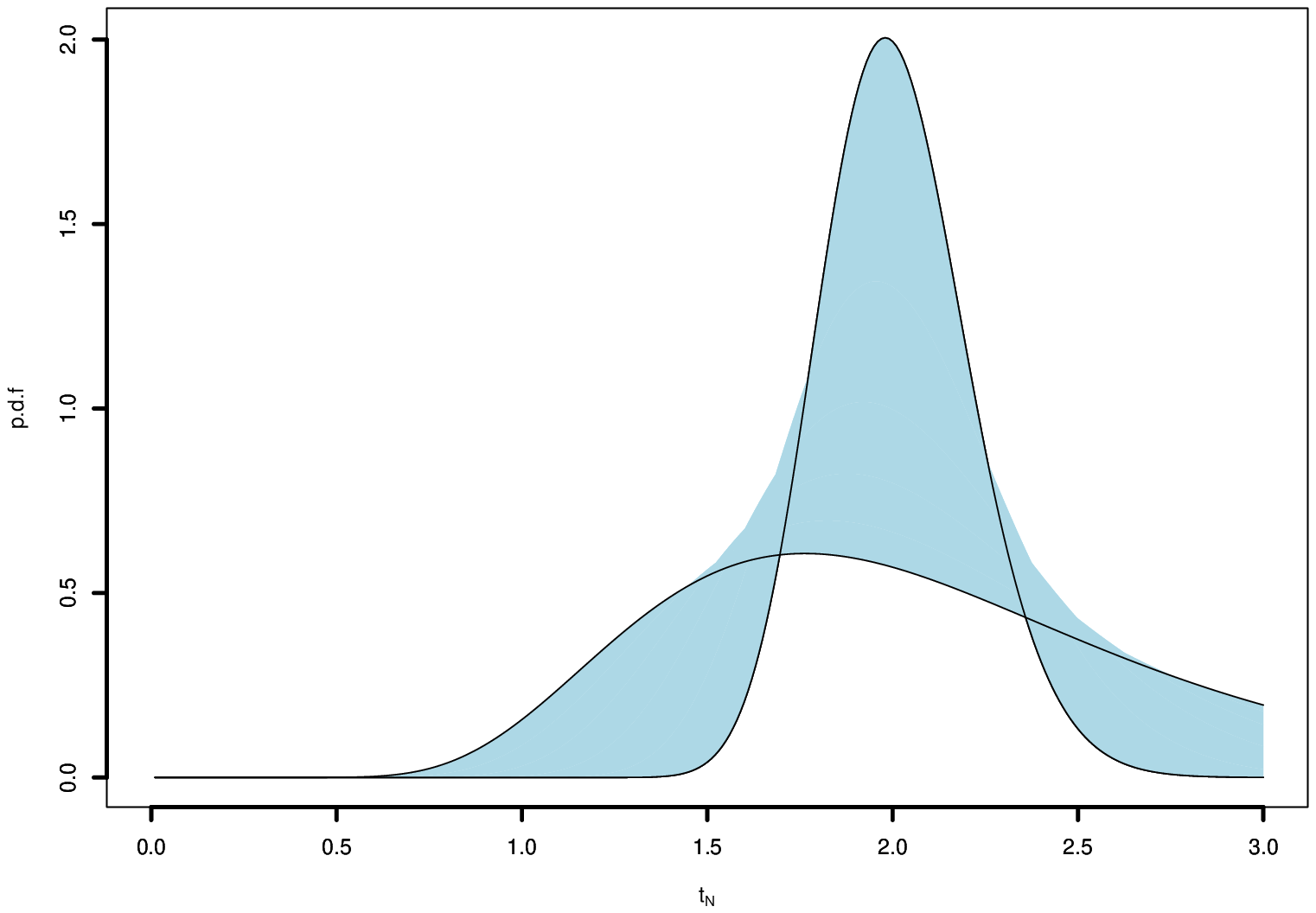}
			\\
			$\beta = 1, \alpha_{\scriptscriptstyle N} \in [0.1, 0.35]$ & 
			$\beta = 2, \alpha_{\scriptscriptstyle N} \in [0.1, 0.35]$  
			\\
			\includegraphics[angle=-90,scale = 0.2]{ 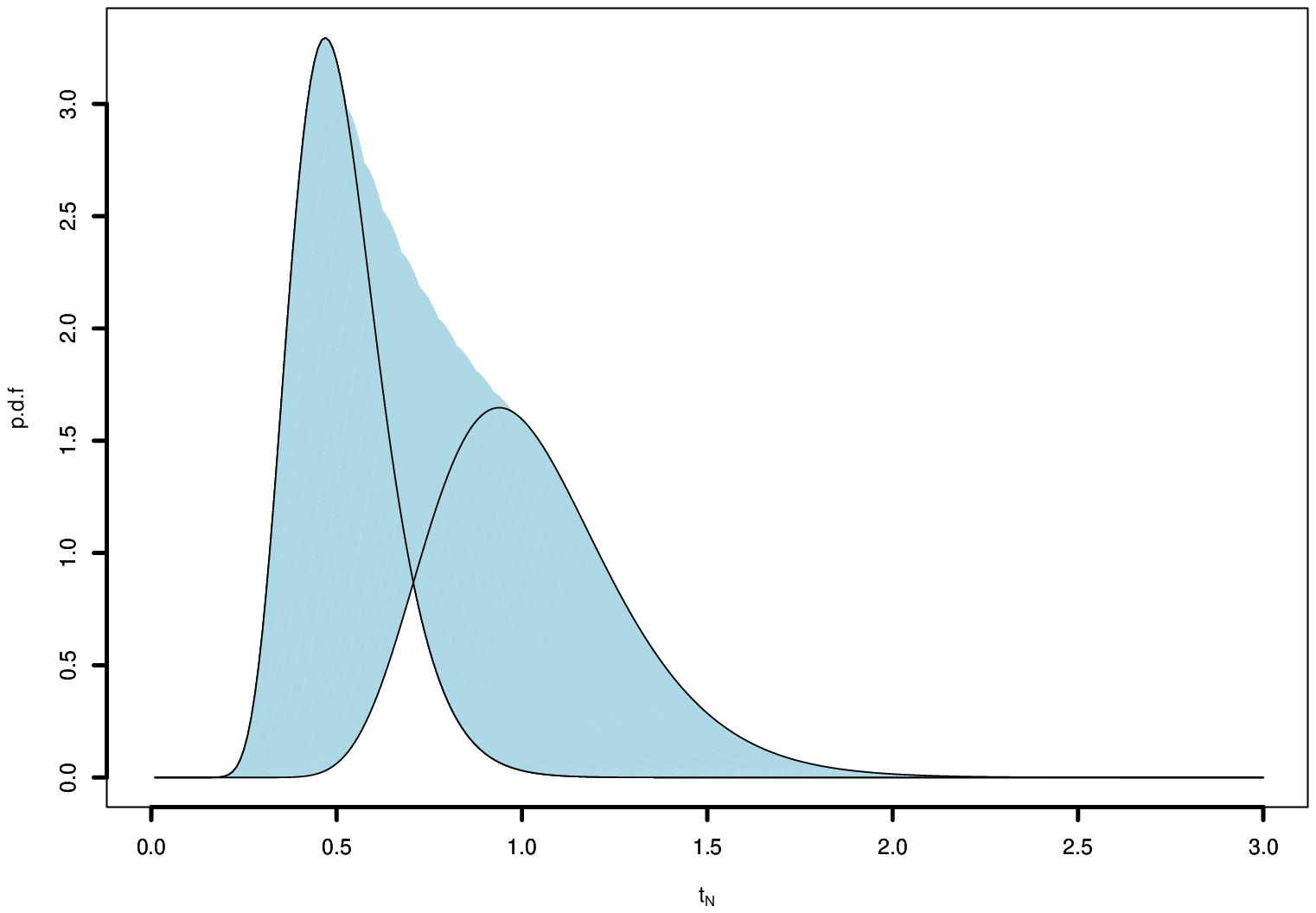}&
			\includegraphics[angle=-90,scale = 0.2]{ 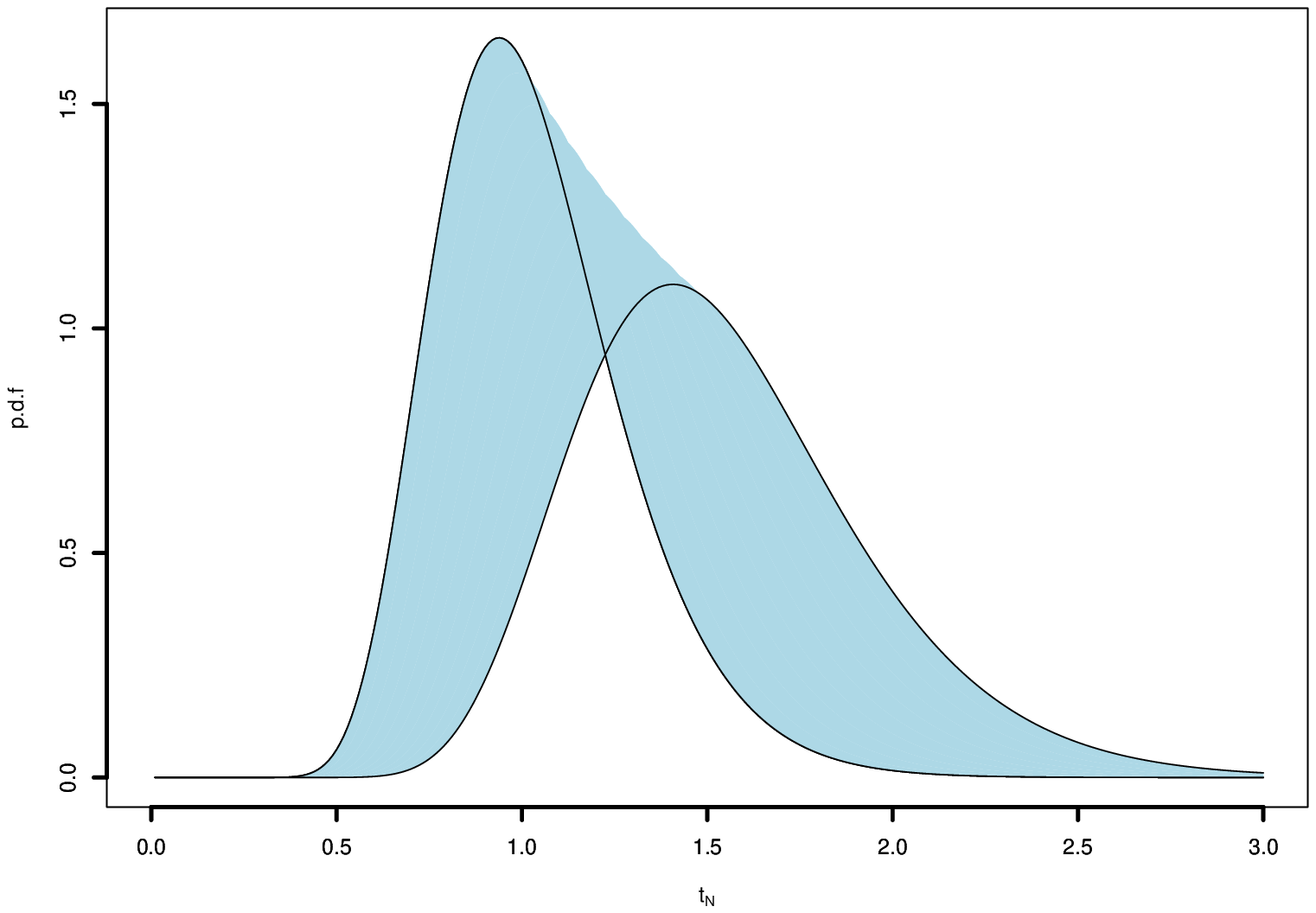}
			\\
			$\alpha = 0.25, \beta_{\scriptscriptstyle N} \in [0.5, 1]$ &
			$\alpha = 0.25, \beta_{\scriptscriptstyle N} \in [1, 1.5]$
			\\
			\includegraphics[angle=-90,scale = 0.2]{ 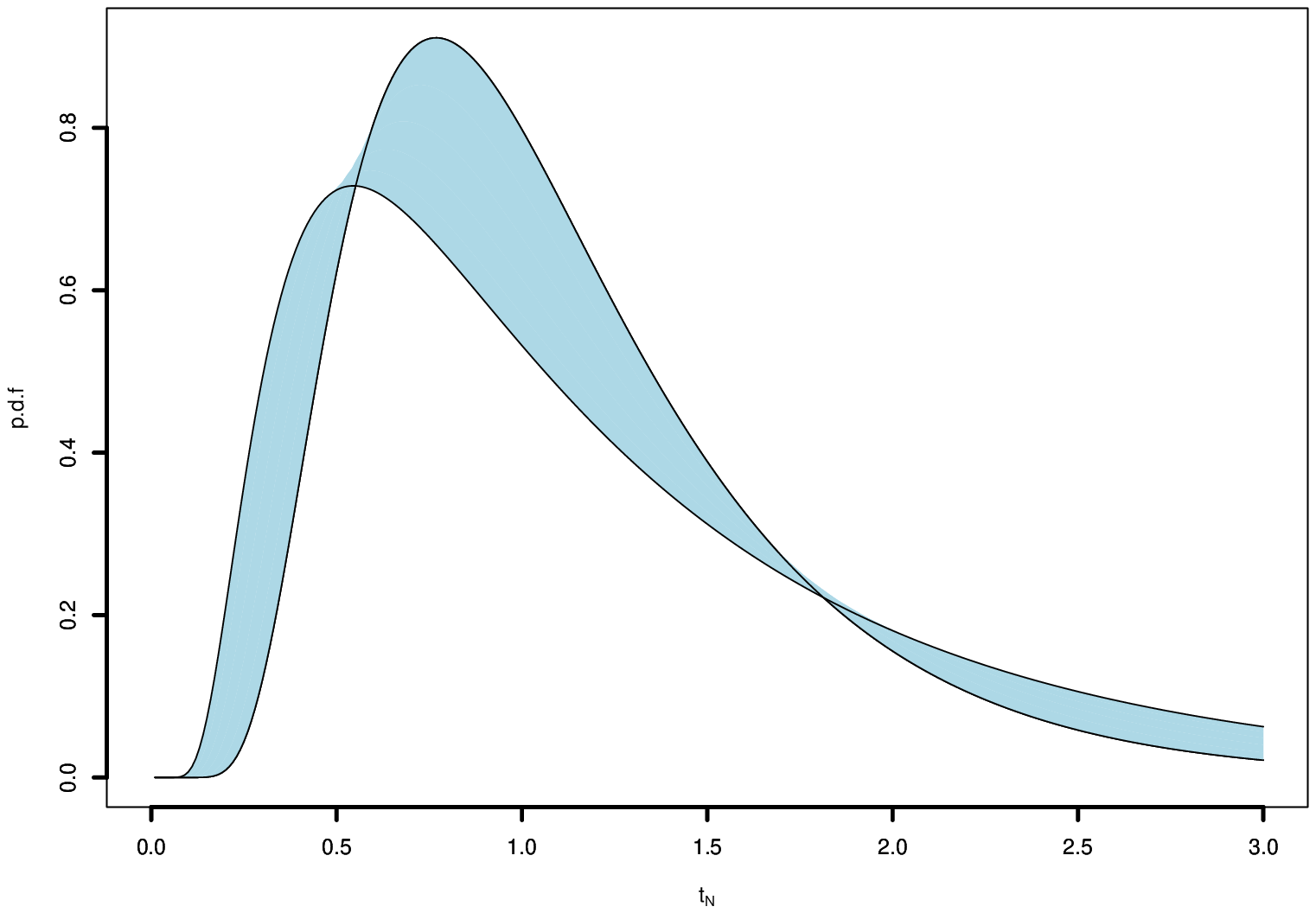} &  
			\includegraphics[angle=-90,scale = 0.2]{ 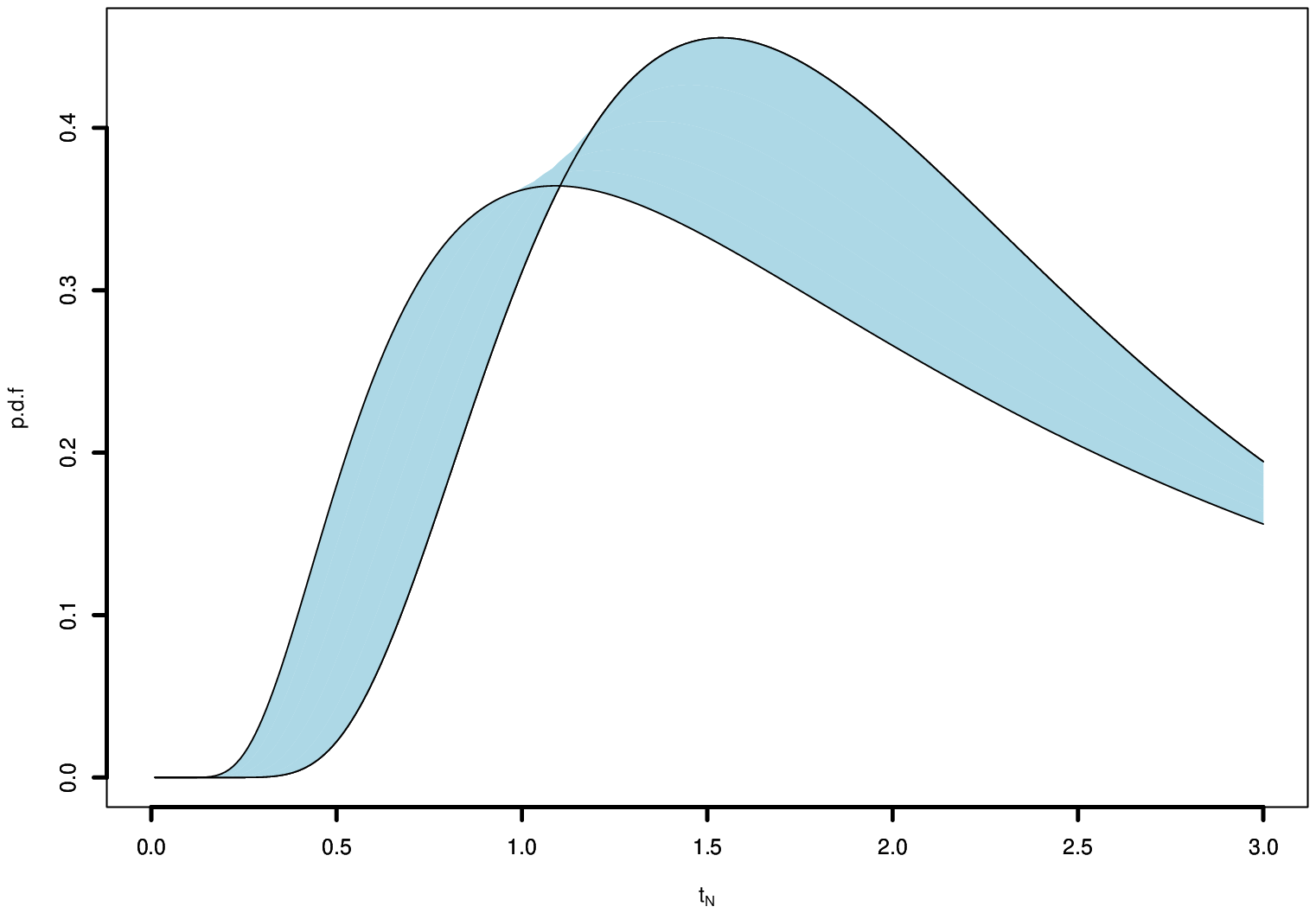}  
			\\
			$\beta = 1, \alpha_{\scriptscriptstyle N} \in [0.5, 0.75]$ & 
			$\beta = 2, \alpha_{\scriptscriptstyle N} \in [0.5, 0.75]$ 
			\\
			\includegraphics[angle=-90,scale = 0.2]{ 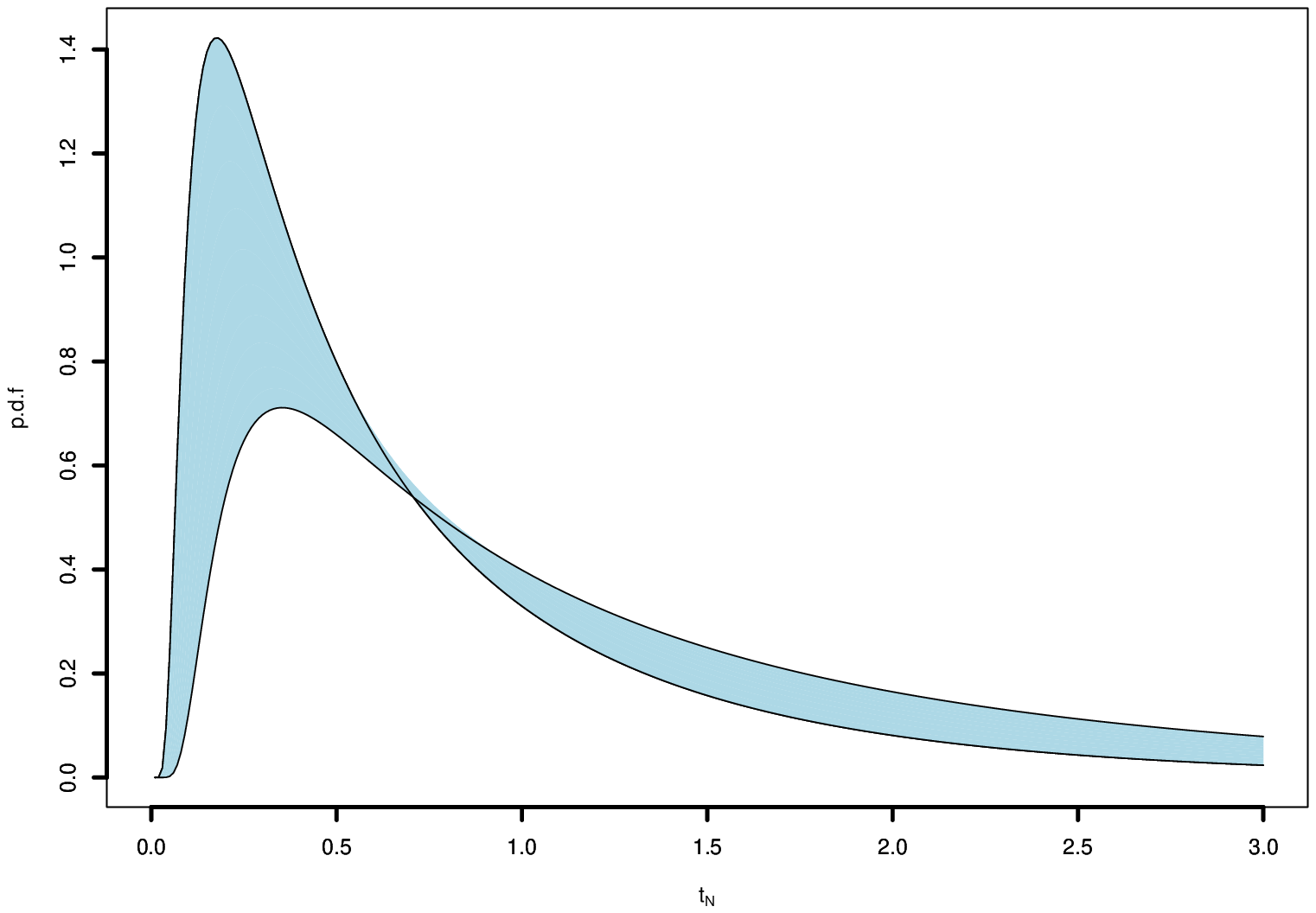}& 
			\includegraphics[angle=-90,scale = 0.2]{ 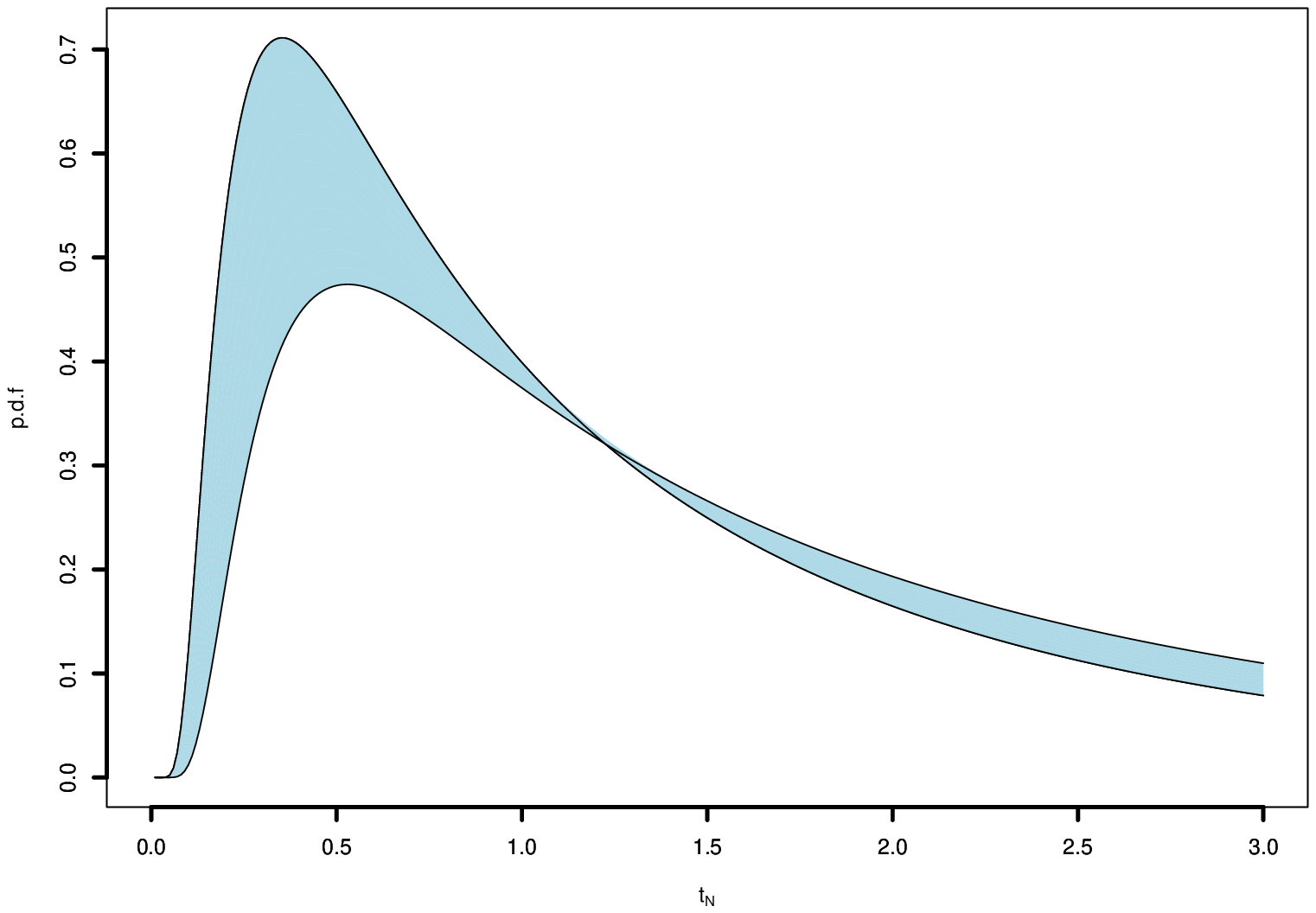}
			\\
			$\alpha = 1, \beta_{\scriptscriptstyle N} \in [0.5, 1]$ &
			$\alpha = 1, \beta_{\scriptscriptstyle N} \in [1, 1.5]$\\
		\end{tabular}
	\end{center}
	\caption{The PDF plots of NBSD against different neutrosophic parameters' values.}\label{fig:pdf}
\end{figure}

\begin{figure}[!htbp]
	 \begin{center}
		 \begin{tabular}{cc}
			     \includegraphics[angle=-90,scale = 0.2]{ 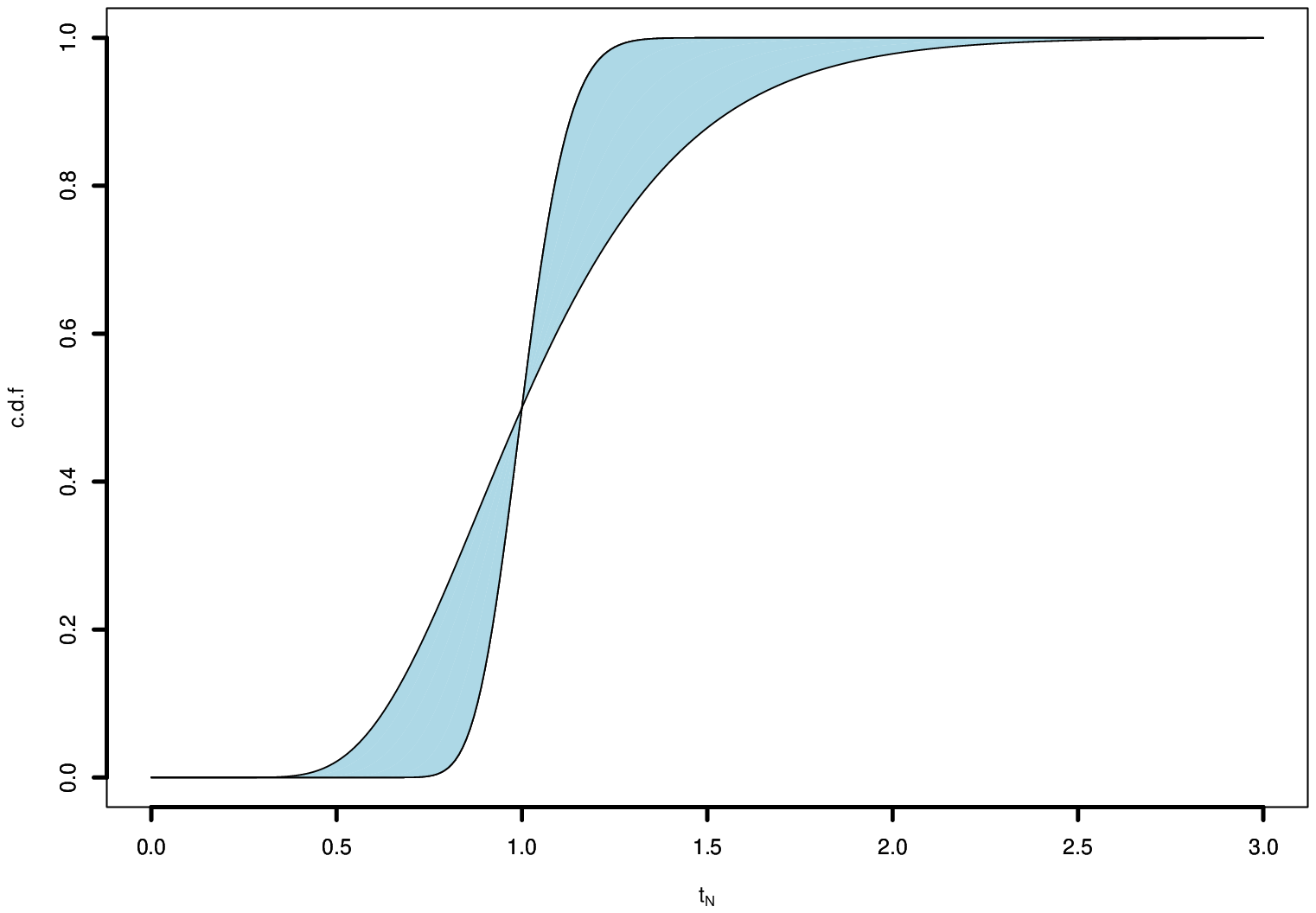}&  
			     \includegraphics[angle=-90,scale = 0.2]{ 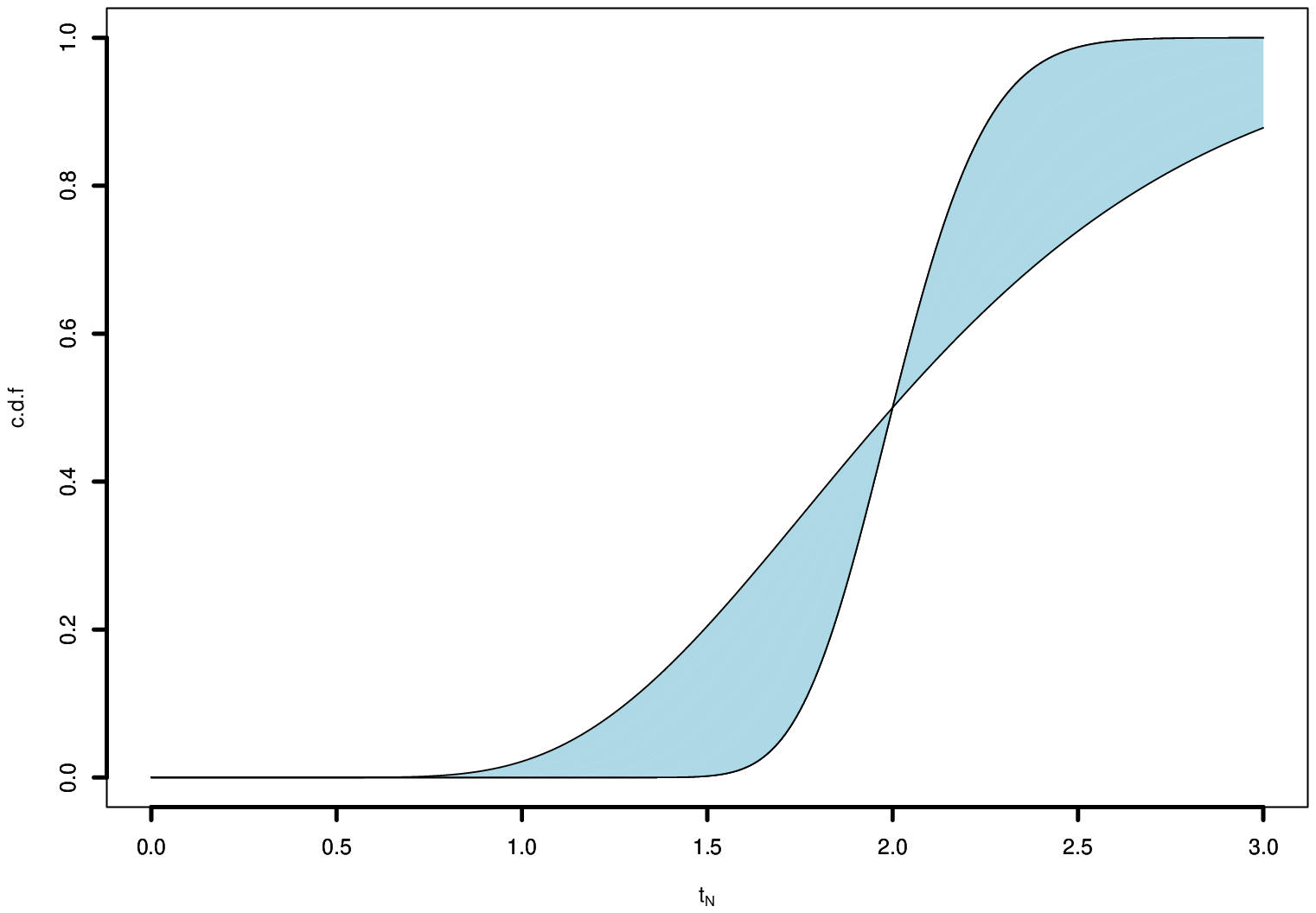}
			     \\
			     $\beta = 1, \alpha_{\scriptscriptstyle N} \in [0.1, 0.35]$ & 
			     $\beta = 2, \alpha_{\scriptscriptstyle N} \in [0.1, 0.35]$ 
			     \\
			     \includegraphics[angle=-90,scale = 0.2]{ 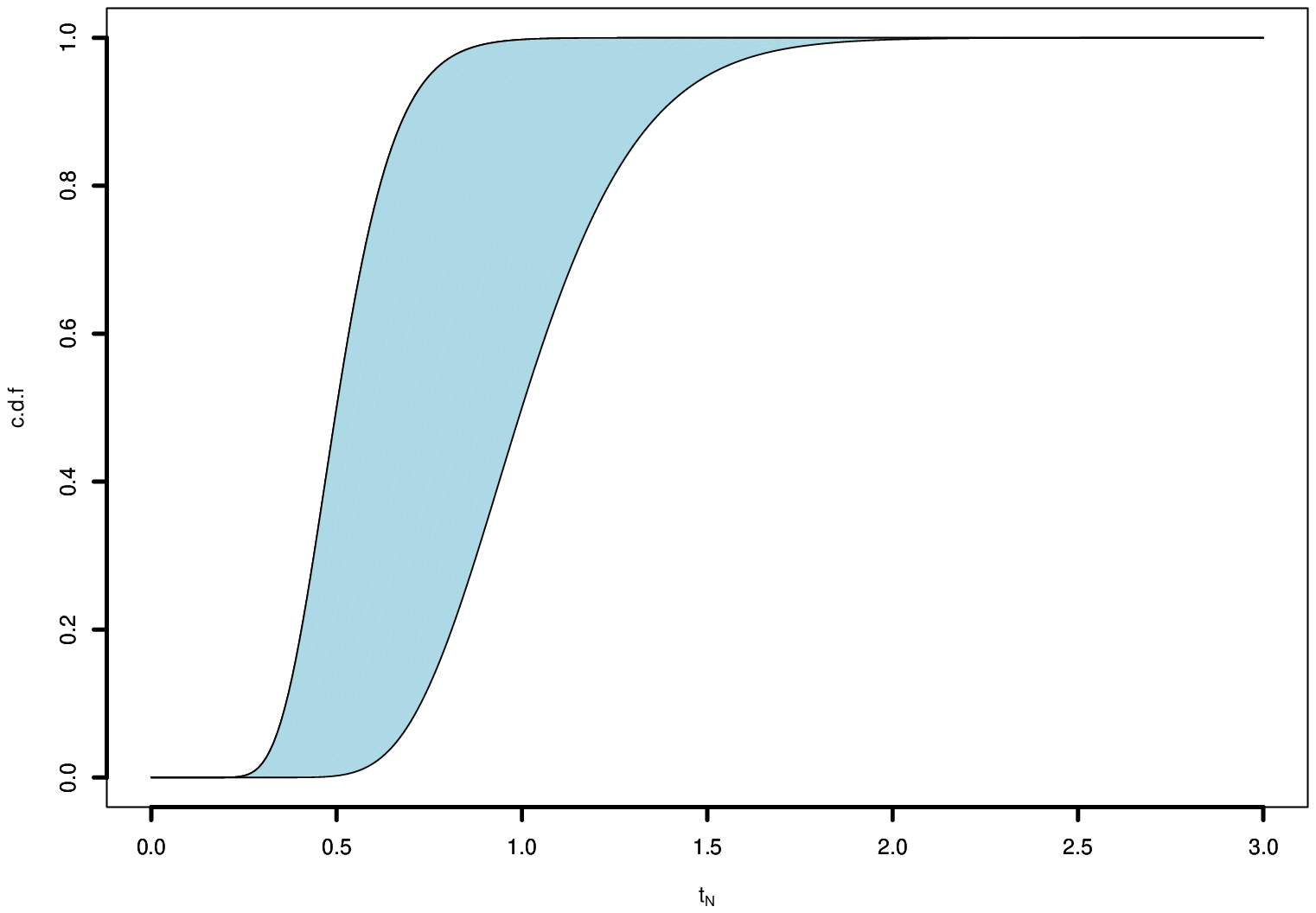}&
			     \includegraphics[angle=-90,scale = 0.2]{ 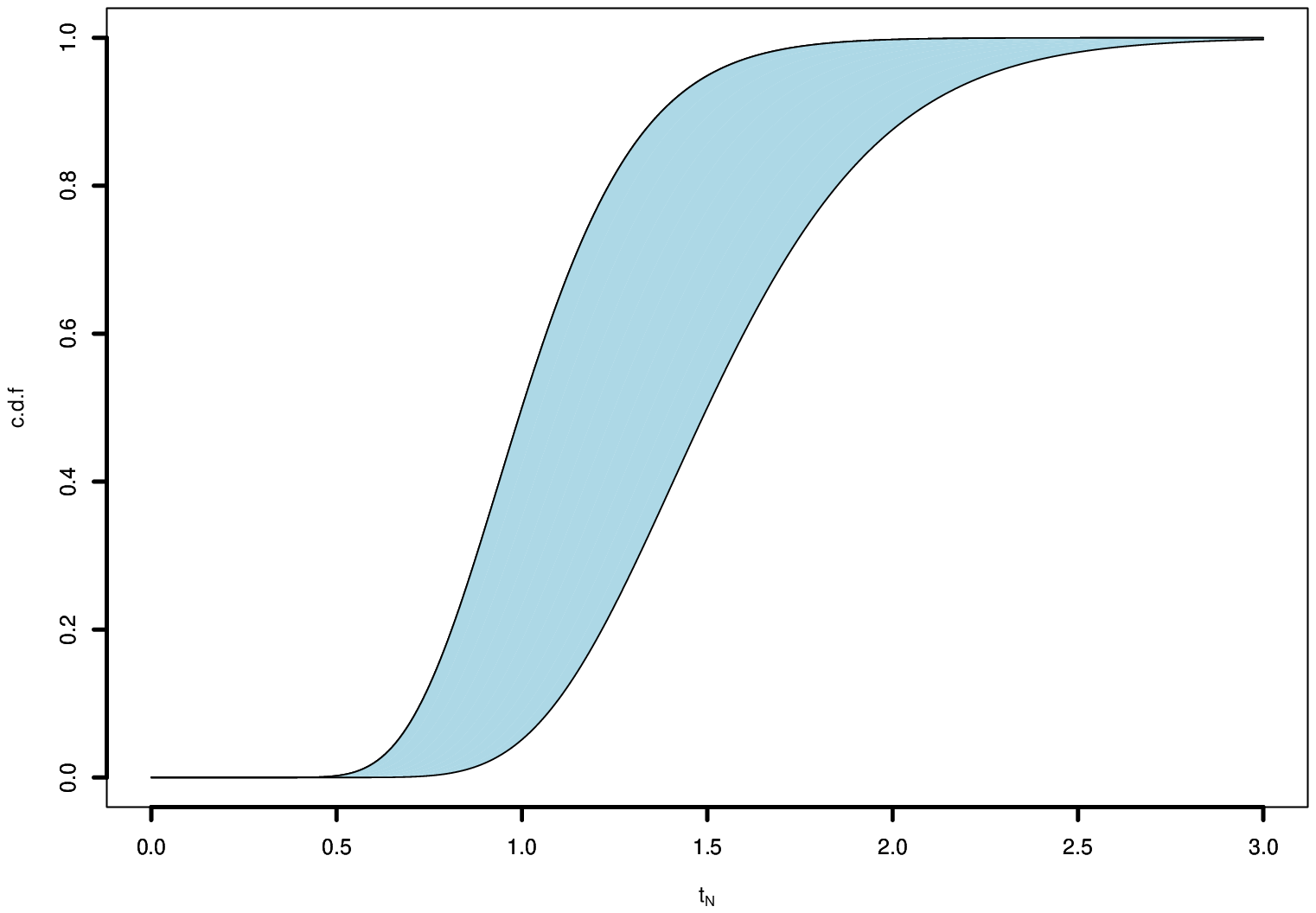}
			     \\
			     $\alpha = 0.25, \beta_{\scriptscriptstyle N} \in [0.5, 1]$ &
			     $\alpha = 0.25, \beta_{\scriptscriptstyle N} \in [1, 1.5]$
			     \\
			     \includegraphics[angle=-90,scale = 0.2]{ 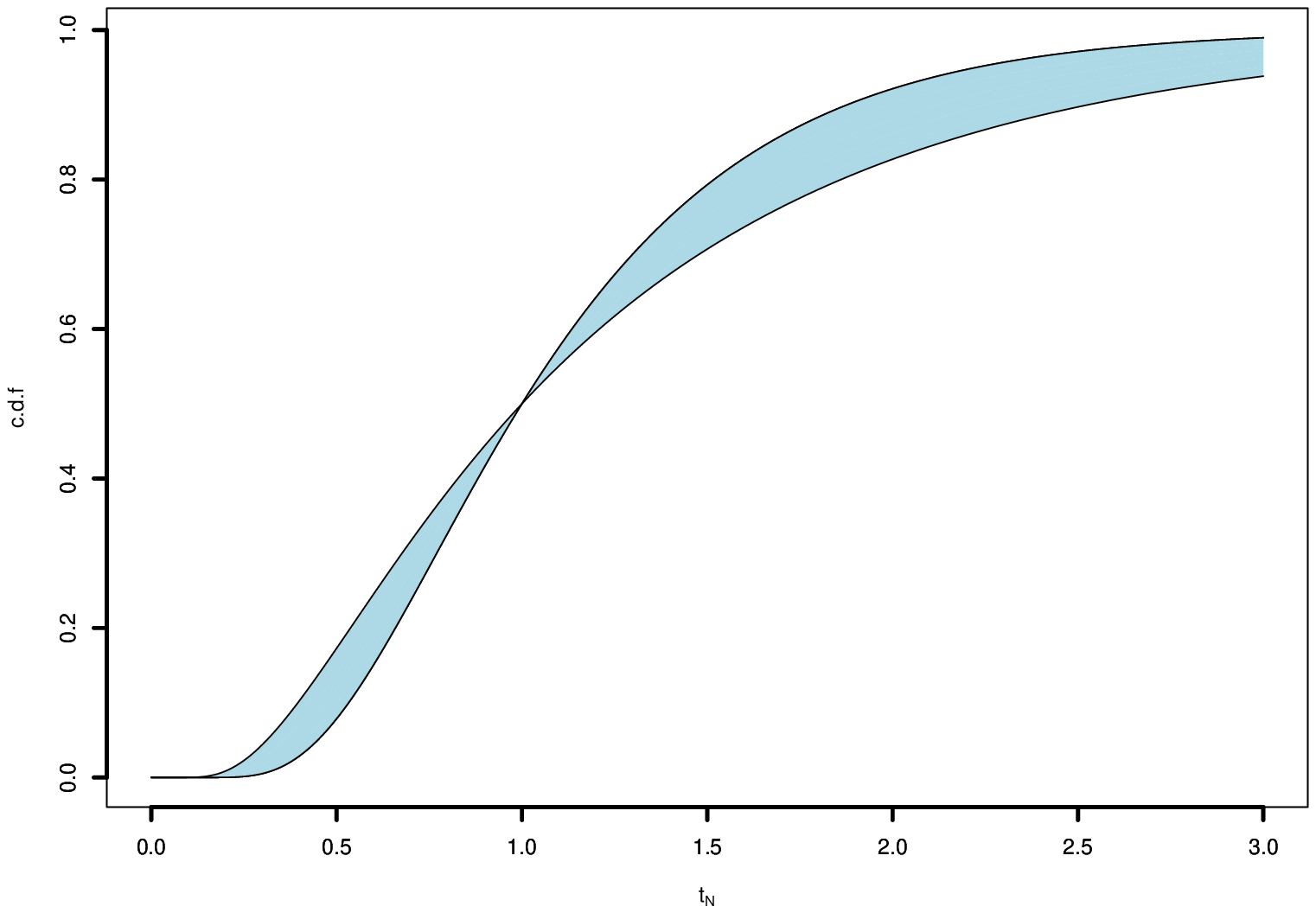} &  
			     \includegraphics[angle=-90,scale = 0.2]{ 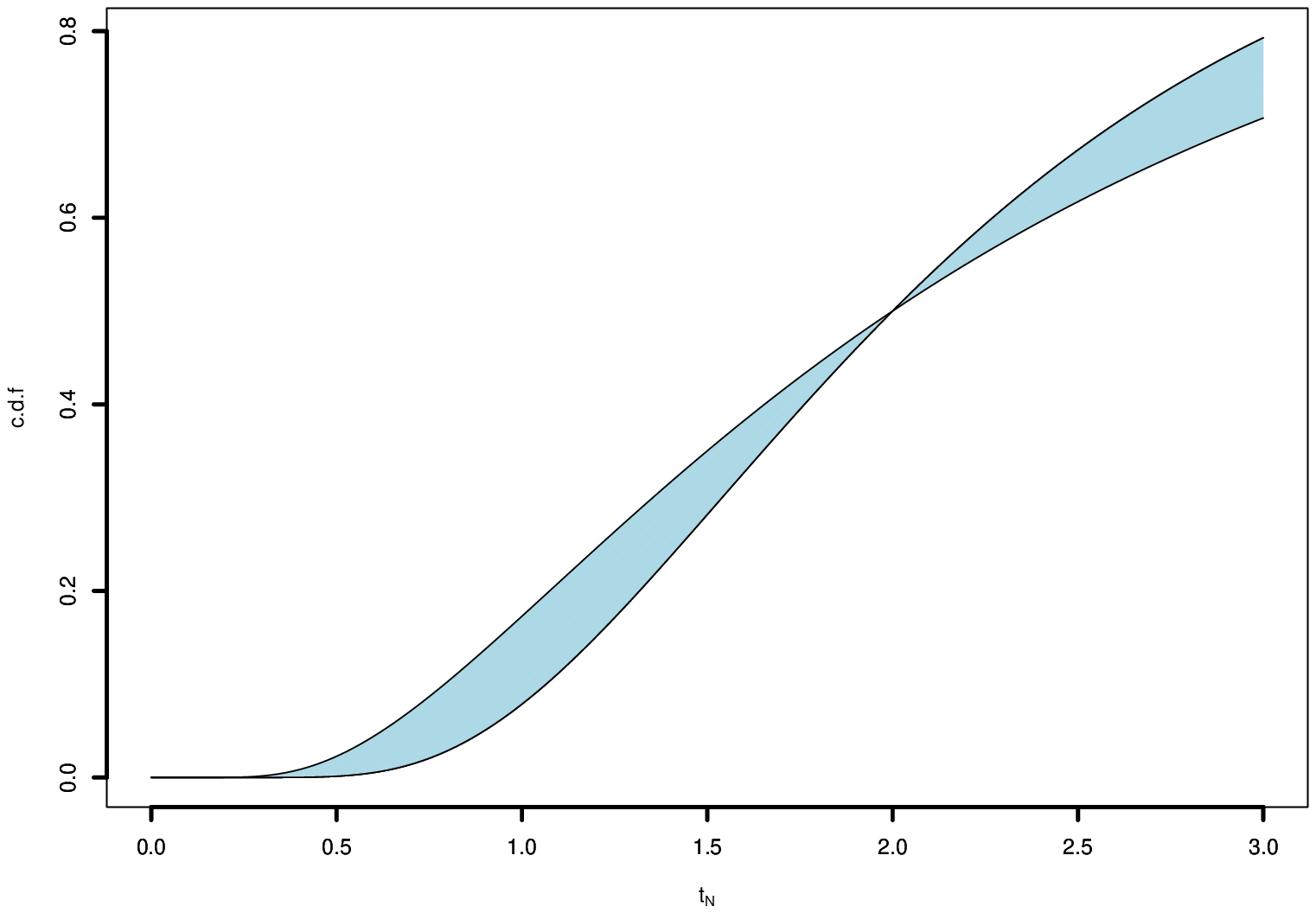} 
			     \\
			     $\beta = 1, \alpha_{\scriptscriptstyle N} \in [0.5, 0.75]$ & 
			     $\beta = 2, \alpha_{\scriptscriptstyle N} \in [0.5, 0.75]$ 
			     \\
			     \includegraphics[angle=-90,scale = 0.2]{ 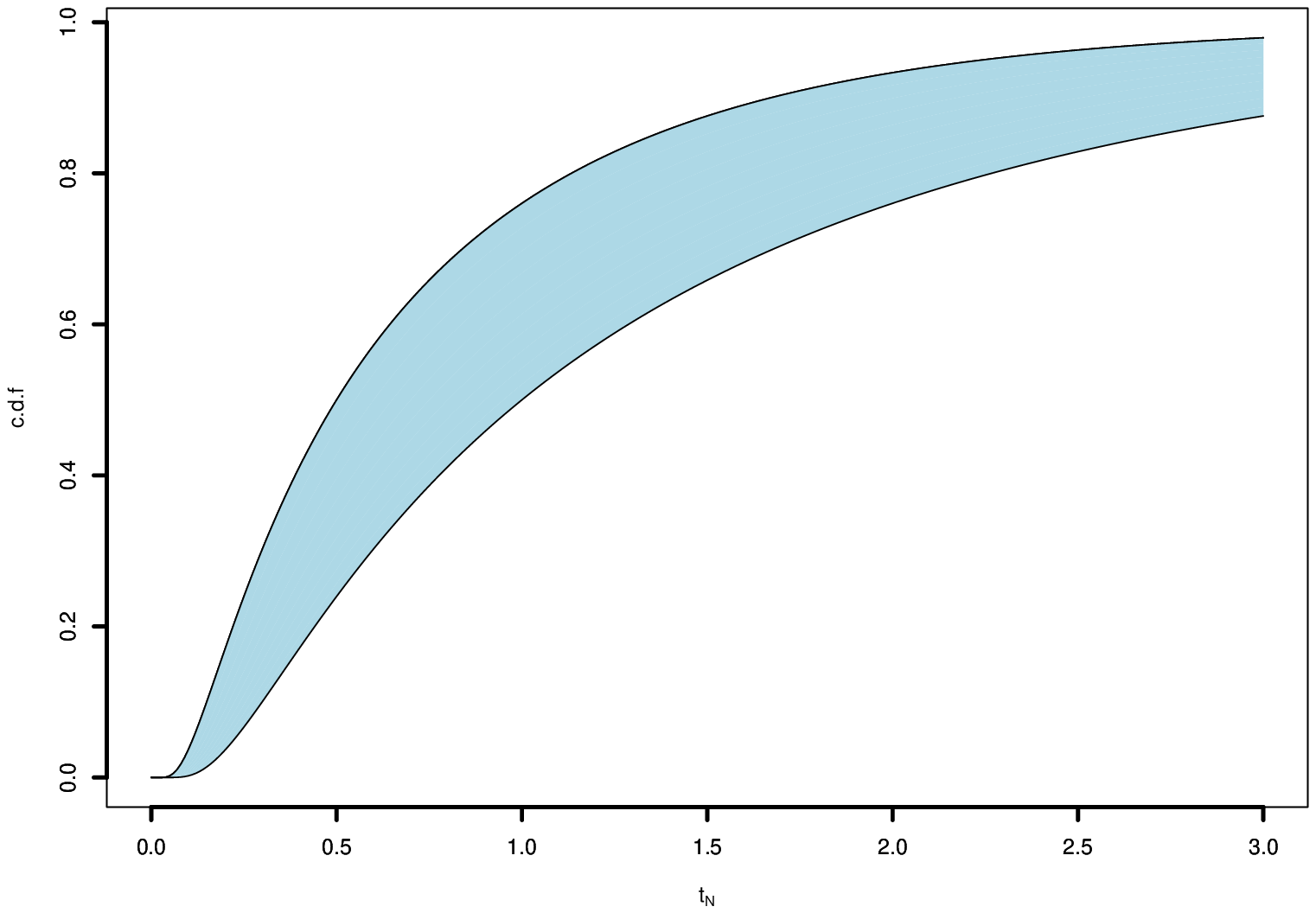}& 
			     \includegraphics[angle=-90,scale = 0.2]{ 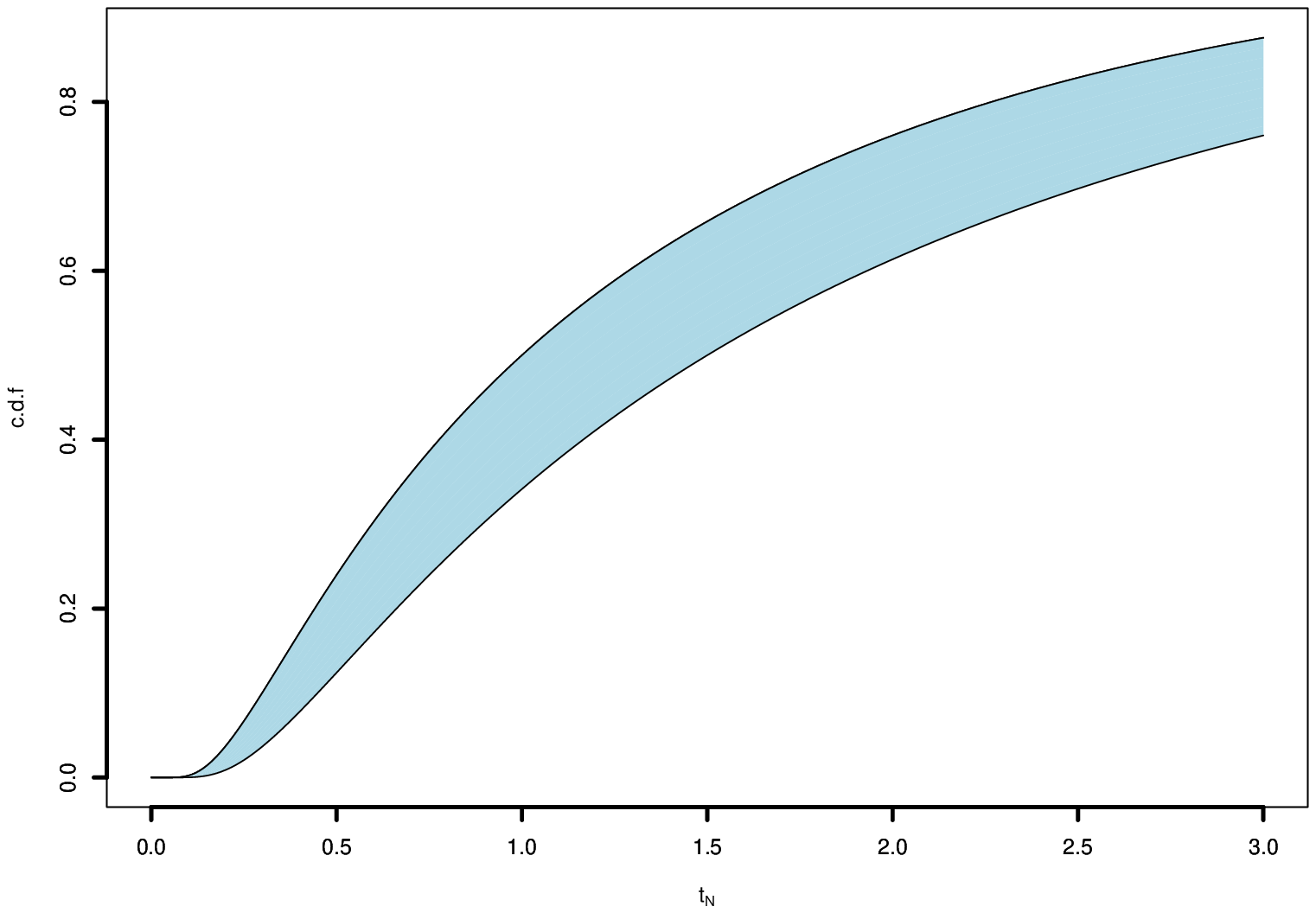}
			     \\
			     $\alpha = 1, \beta_{\scriptscriptstyle N} \in [0.5, 1]$ &
			     $\alpha = 1, \beta_{\scriptscriptstyle N} \in [1, 1.5]$
			     \\
			 \end{tabular}
		 \end{center}
	     \caption{The CDF plots of NBSD against different neutrosophic parameters' values.}\label{fig:cdf}
	 \end{figure}

\begin{figure}[!htbp]
	 \begin{center}
		 \begin{tabular}{cc}
			 \includegraphics[angle=-90,scale = 0.2]{ 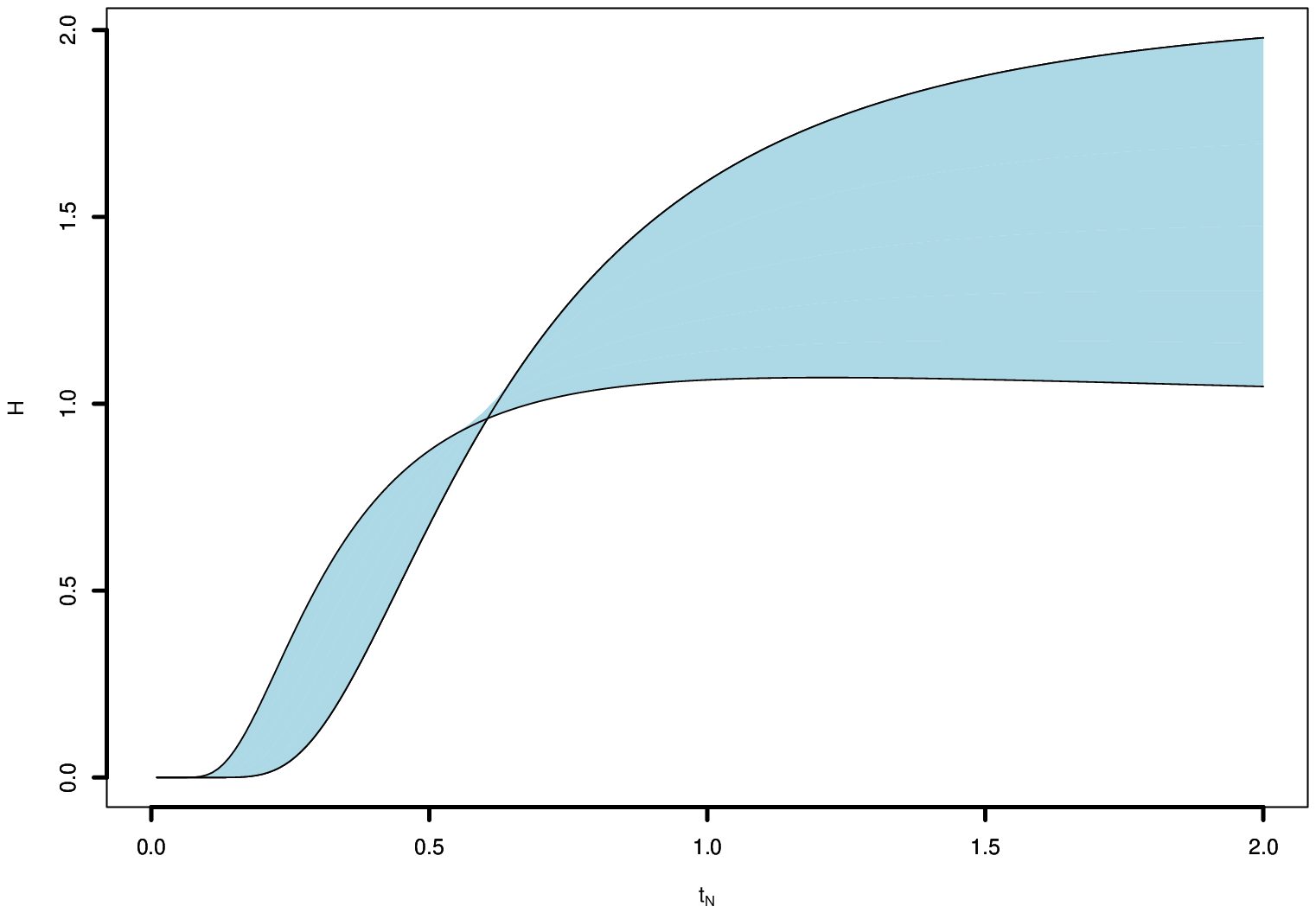} &  
			 \includegraphics[angle=-90,scale = 0.2]{ 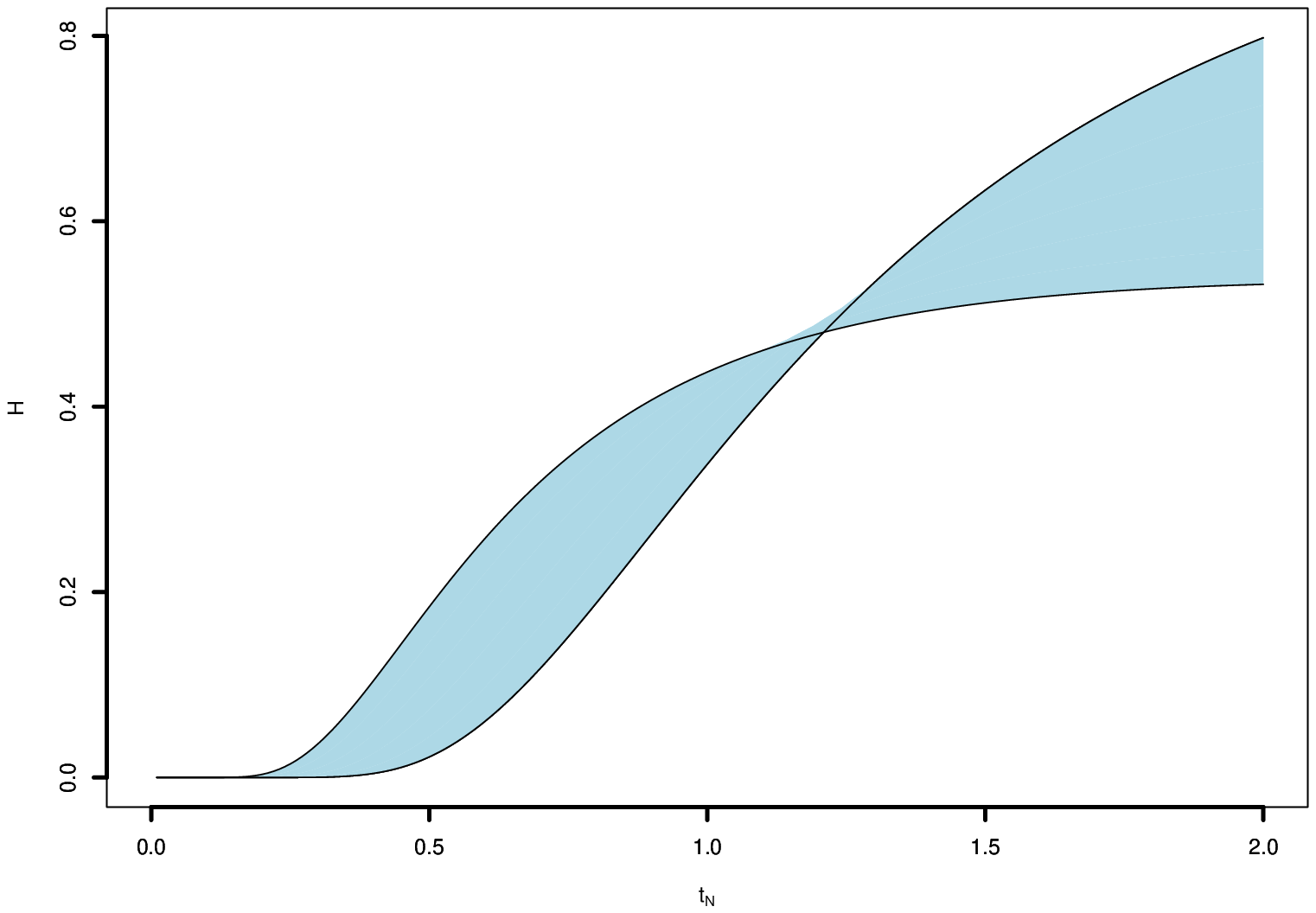}  \\
			 $\beta = 1, \alpha_{\scriptscriptstyle N} \in [0.5, 0.75]$ & 
			 $\beta = 2, \alpha_{\scriptscriptstyle N} \in [0.5, 0.75]$ \\
			 \includegraphics[angle=-90,scale = 0.2]{ 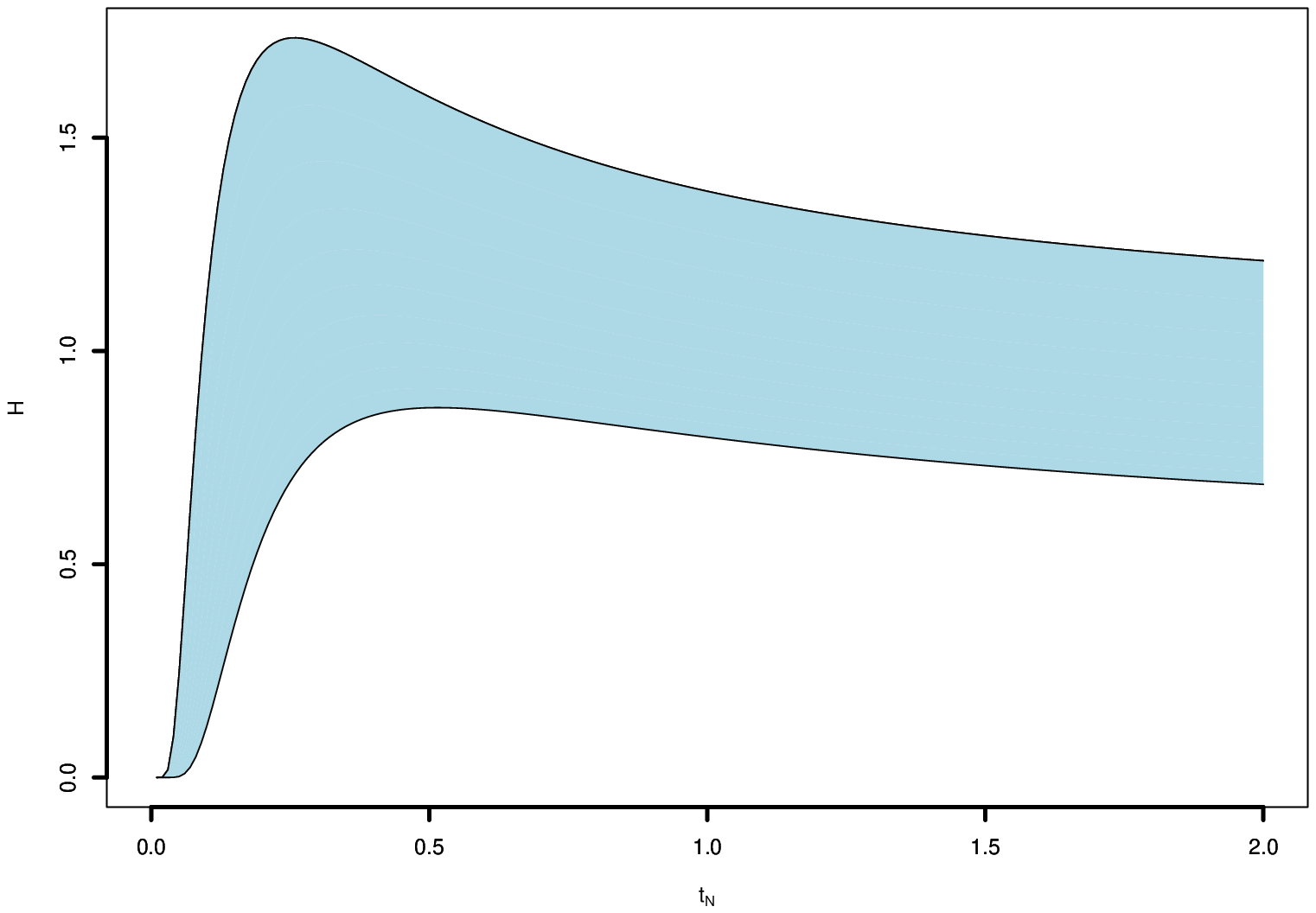}& 
			 \includegraphics[angle=-90,scale = 0.2]{ 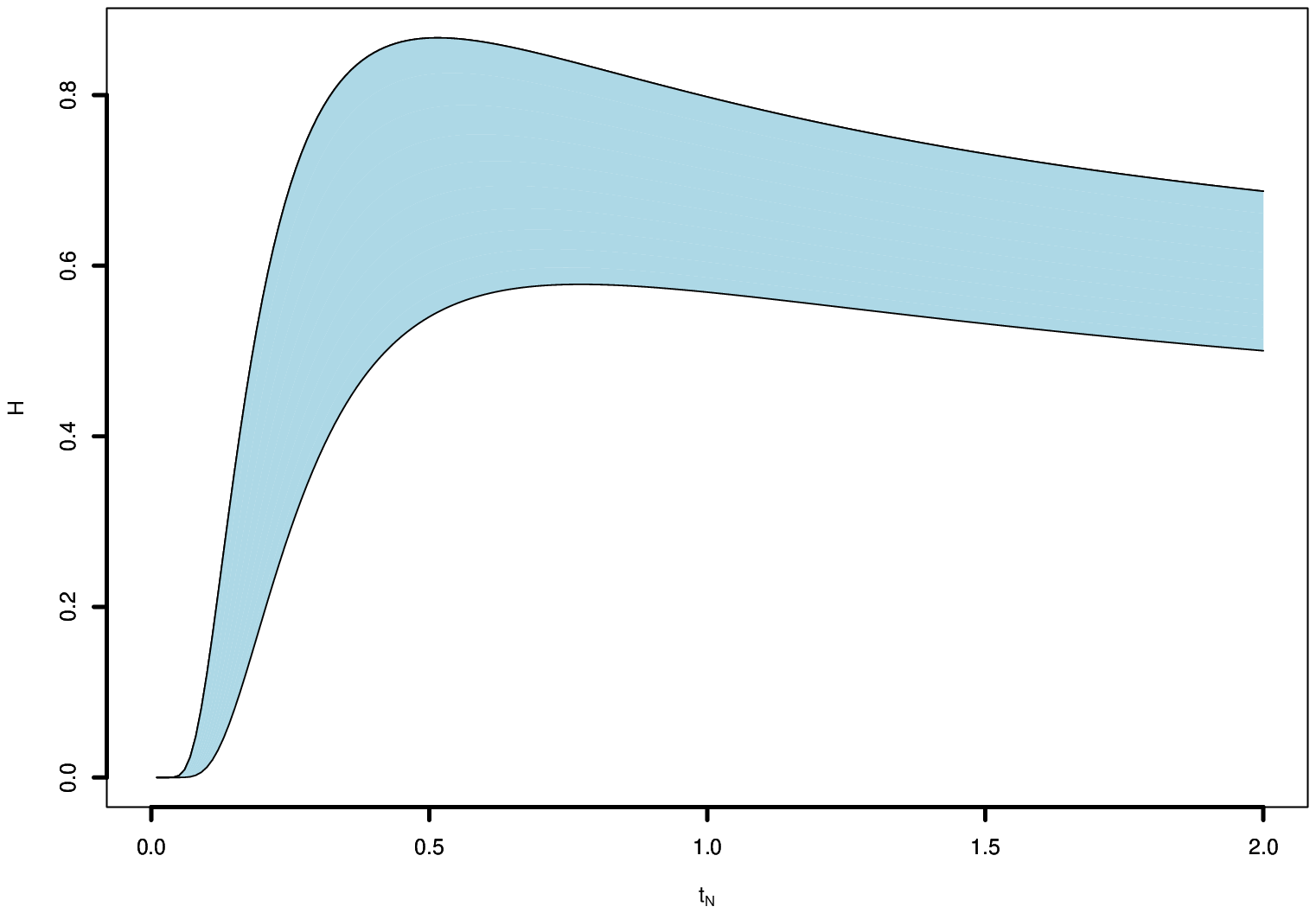}\\
			 $\alpha = 1, \beta_{\scriptscriptstyle N} \in [0.5, 1]$ &
			 $\alpha = 1, \beta_{\scriptscriptstyle N} \in [1, 1.5]$\\
			 \includegraphics[angle=-90,scale = 0.2]{ 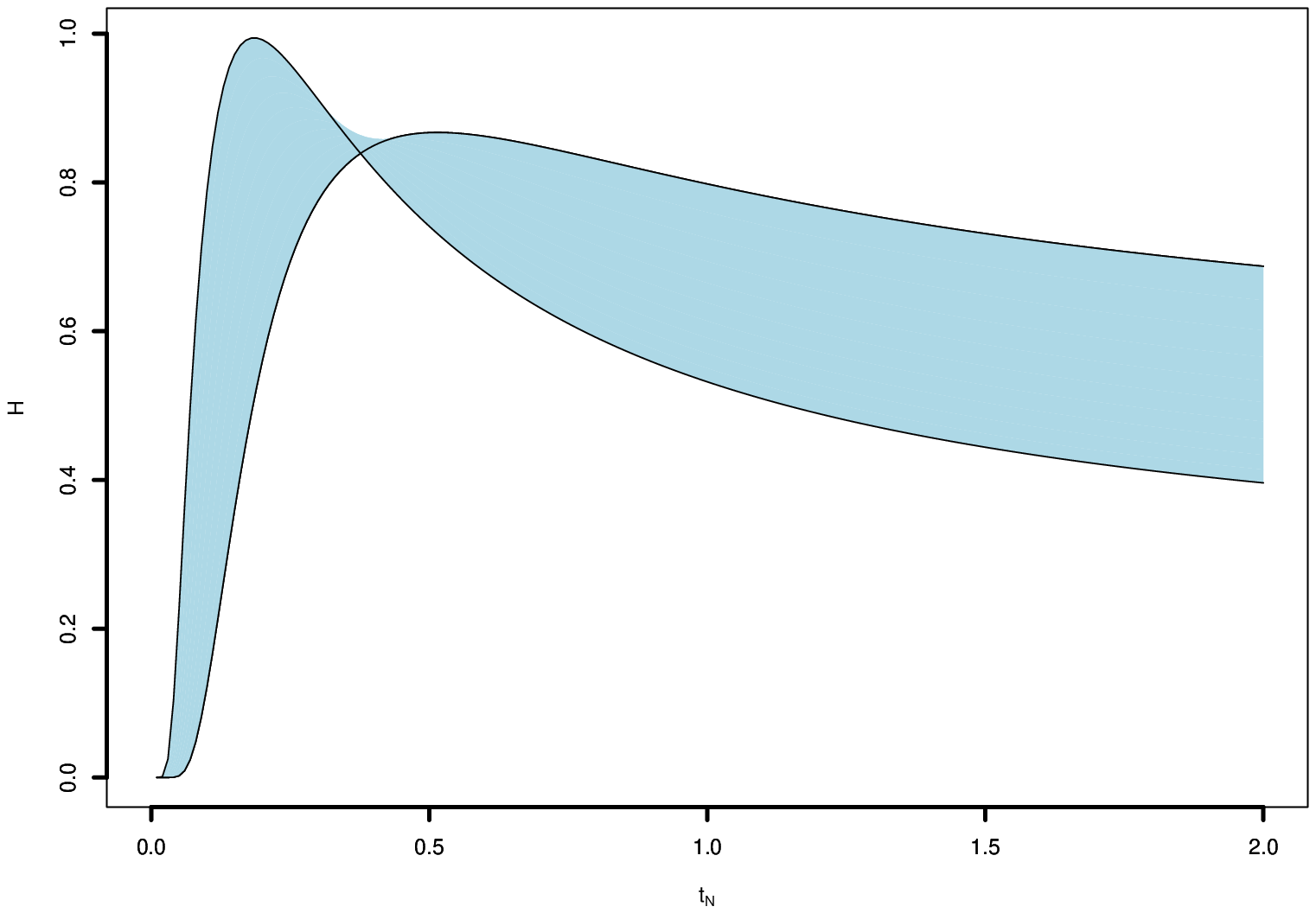} &  
			 \includegraphics[angle=-90,scale = 0.2]{ 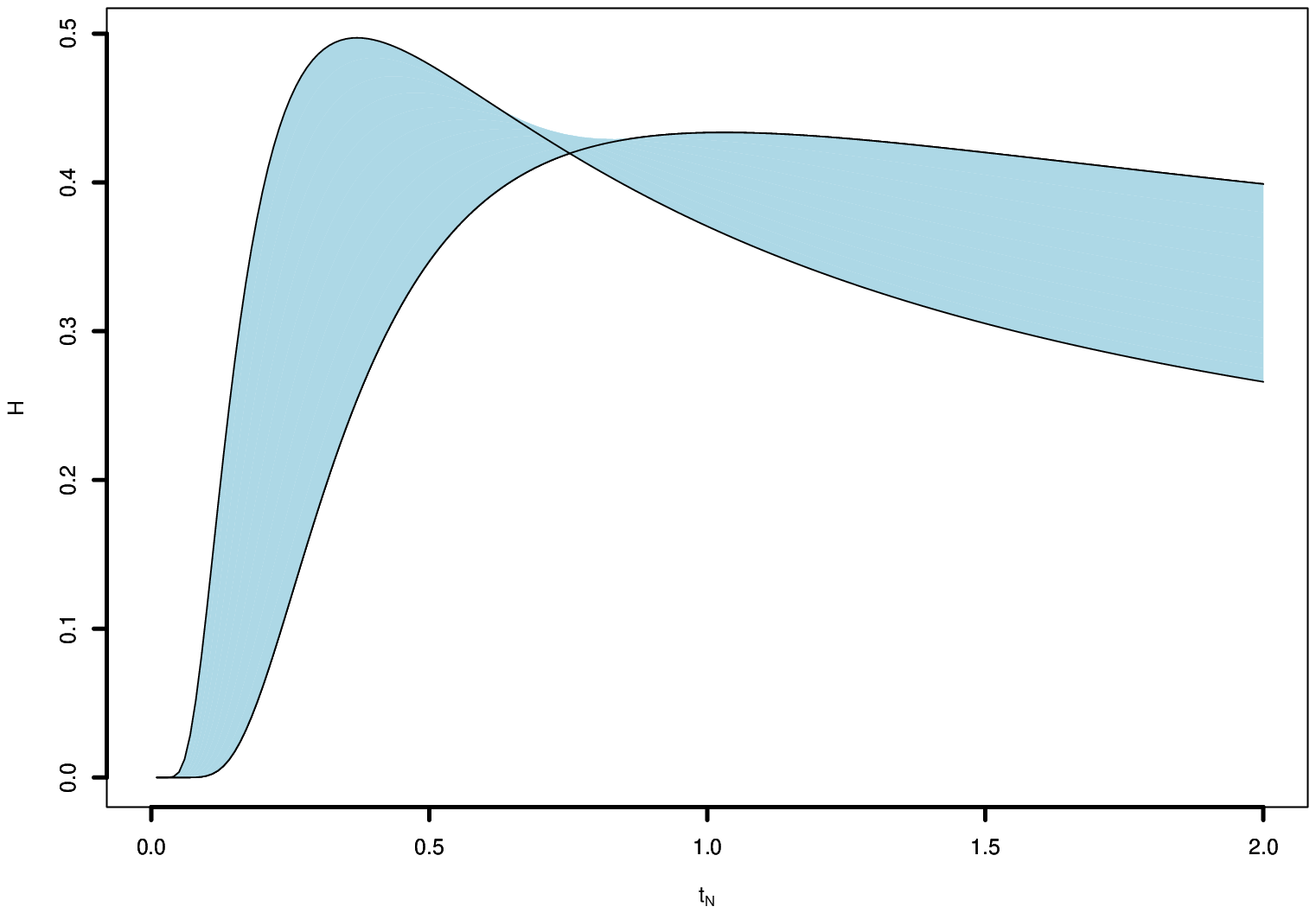} \\
			 $\beta = 1, \alpha_{\scriptscriptstyle N} \in [1, 1.5]$ &
			 $\beta = 2, \alpha_{\scriptscriptstyle N} \in [1, 1.5]$ \\
			 \includegraphics[angle=-90,scale = 0.2]{ 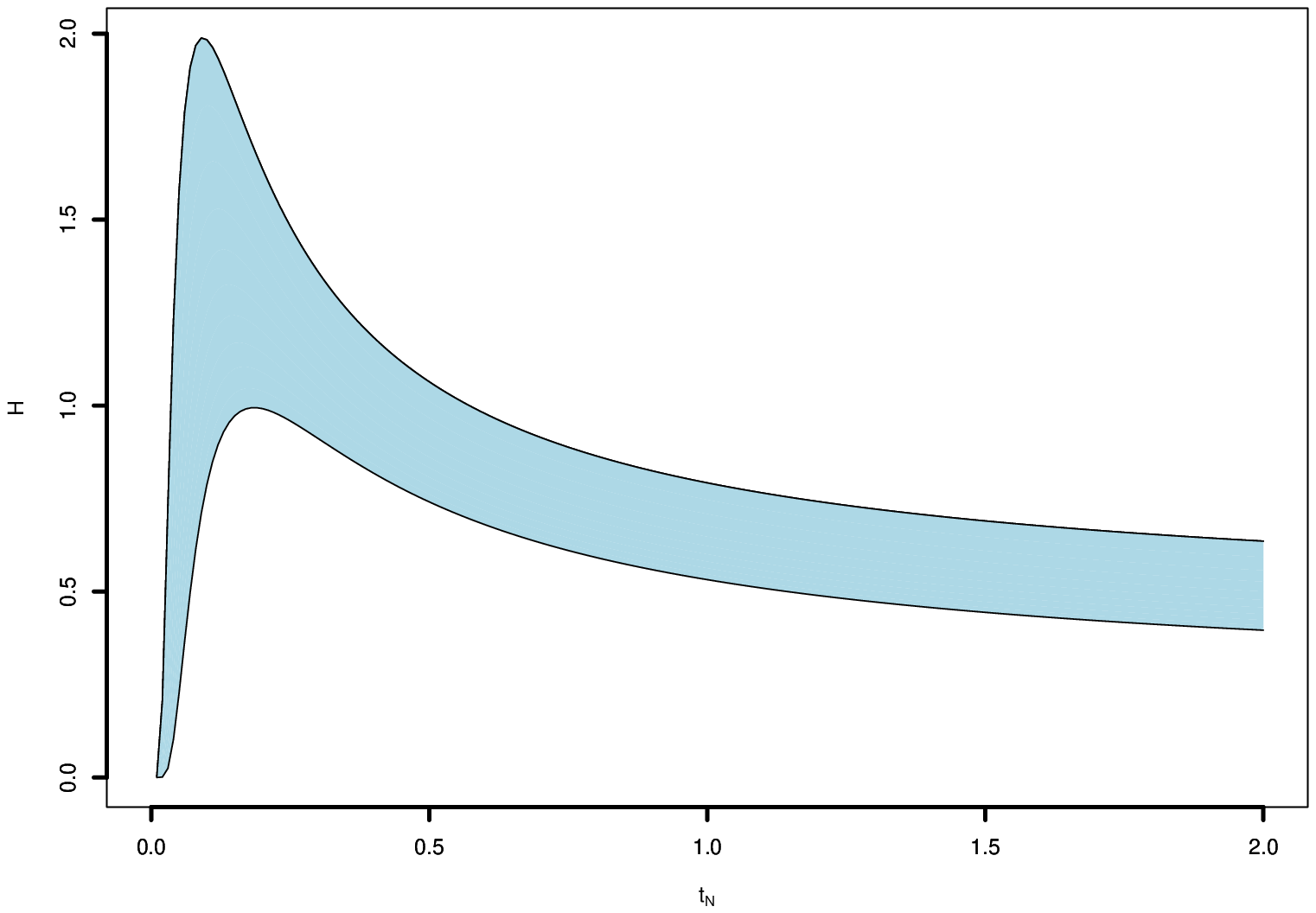}& 
			 \includegraphics[angle=-90,scale = 0.2]{ 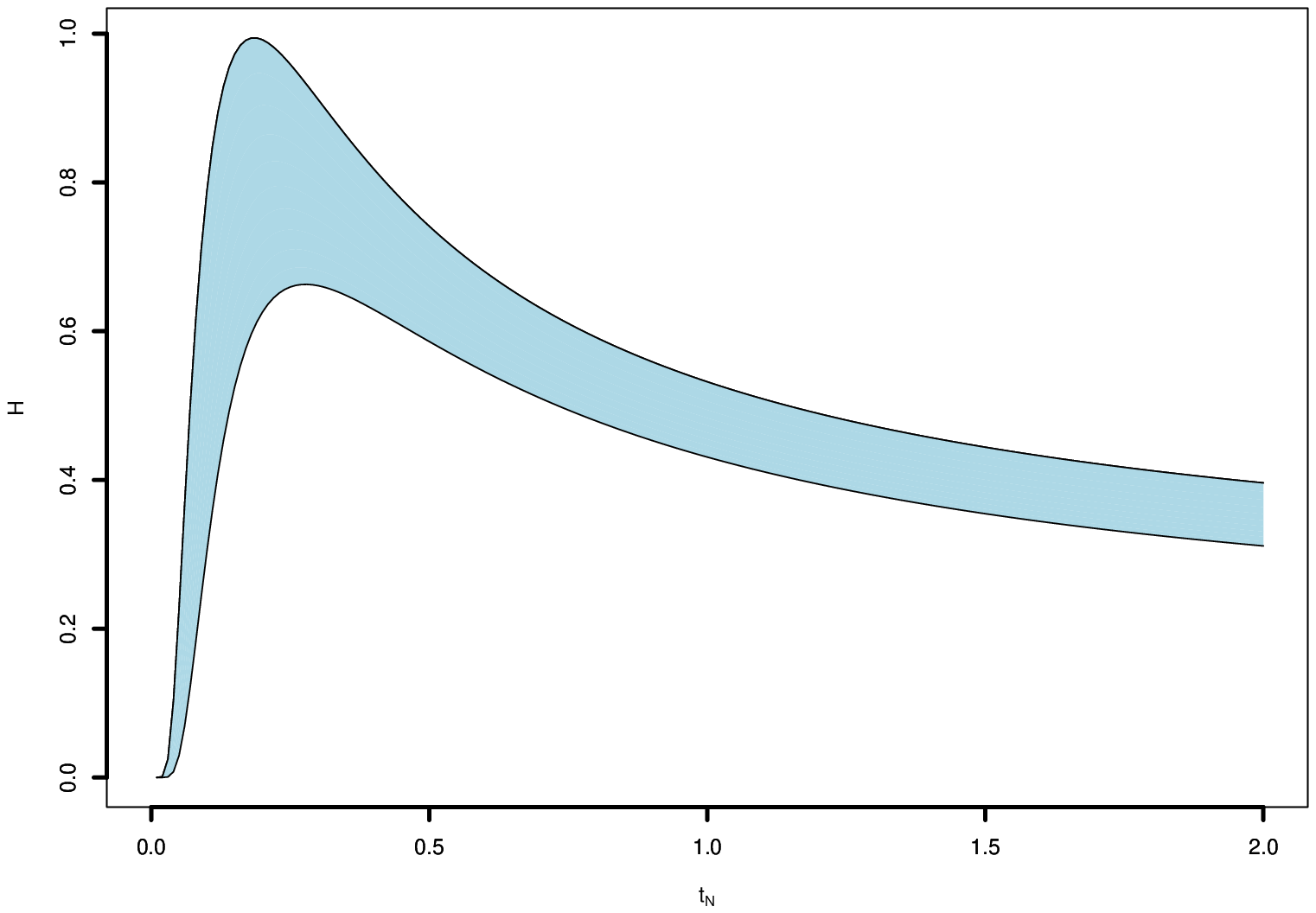}\\
			 $\alpha = 1.5, \beta_{\scriptscriptstyle N} \in [0.5, 1]$ &
			 $\alpha = 1.5, \beta_{\scriptscriptstyle N} \in [1, 1.5]$ \\
			 \end{tabular}
		 \end{center}
	     \caption{The hazard function plots of NBSD against different neutrosophic parameters' values.}\label{fig:H}
	 \end{figure}

Figure \ref{fig:pdf} illustrates how PDF values of NBSD change with different neutrosophic scale and/or shape parameters. Figures \ref{fig:cdf} and \ref{fig:H} display the CDF and hazard functions of the NBSD, respectively. Black curves in all plots show the function values in lower and upper boundary of neutrosophic parameters, and the coloured areas represent the function values as parameters change through indeterminacy intervals.

\section{Properties of the NBSD}\label{sec:3}
Some key statistical properties, central tendency, variability, shape, percentiles, moments and correlation are given below.
When $T_{\scriptscriptstyle N} \sim NBS(\alpha_{\scriptscriptstyle N}, \beta_{\scriptscriptstyle N})$, by using the relation in (\ref{eq:1}), and different moments of the standard normal random variable, the central moments of $T_{\scriptscriptstyle N}$, for integer $r$, are

\begin{equation}\label{CMoments}
   \mathrm{E}\left({T_{\scriptscriptstyle N}}^r\right)={\beta_{\scriptscriptstyle N}}^r \sum_{j=0}^r\left(\begin{array}{l}
 2 r \\
 2 j
 \end{array}\right) \sum_{i=0}^j\left(\begin{array}{l}
 i \\
 j
 \end{array}\right) \frac{(2 r-2 j+2 i) !}{2^{r-j+i}(r-j+i) !}\left(\frac{\alpha_{\scriptscriptstyle N}}{2}\right)^{2 r-2 j+2 i}.  
\end{equation}

 All the moments can be obtained using central moments from \eqref{CMoments}.  The most important statistical characteristics are summarized in Table \ref{SC}. 
\begin{table}[!htbp]
\caption{The most important statistical characteristics of NBSD.}\label{SC}
\centering
\renewcommand{\arraystretch}{1.7}
\begin{tabular}{ll}
\hline
Mean (E) & $E(T_{\scriptscriptstyle N})=\beta_{\scriptscriptstyle N}\left(1+\frac{{\alpha _{\scriptscriptstyle N}}^2}{2}\right)$\\
\hline
Variance (V) &
$\operatorname{Var}(T_{\scriptscriptstyle N})=(\alpha_{\scriptscriptstyle N} {\beta_{\scriptscriptstyle N}})^2\left(1+\frac{5}{4} {\alpha_{\scriptscriptstyle N}}^2\right)$\\
\hline
Coefficient of variation (CV) & $CV(T_{\scriptscriptstyle N})=\frac{\sqrt{5 \alpha_{\scriptscriptstyle N}^4+4 \alpha_{\scriptscriptstyle N}^2}}{\alpha_{\scriptscriptstyle N}^2+2}$\\
\hline
Skewness (Sk) & 
$Sk(T_{\scriptscriptstyle N})=\frac{16 {\alpha_{\scriptscriptstyle N}}^2\left(11 {\alpha_{\scriptscriptstyle N}}^2+6\right)}{\left(5 {\alpha_{\scriptscriptstyle N}}^2+4\right)^3}$\\
\hline
Kurtosis (Ku)
&
$Ku(T_{\scriptscriptstyle N})=3+\frac{6 {\alpha_{\scriptscriptstyle N}}^2\left(93 {\alpha_{\scriptscriptstyle N}}^2+40\right)}{\left(5 {\alpha_{\scriptscriptstyle N}}^2+4\right)^2}$\\
\hline
\end{tabular}
\end{table}

The median of the distribution is equal to $\beta_{\scriptscriptstyle N}$.
It is clear that as $\alpha_{\scriptscriptstyle N} \rightarrow 0$, the coefficient of kurtosis converges to 3, and the behavior of the NBSD is almost neutrosophic normal with mean and variance around $\beta_{\scriptscriptstyle N}$ and $\beta_{\scriptscriptstyle N}^2 \alpha_{\scriptscriptstyle N}^2$, respectively. 

The descriptive characteristics of the model for various neutrosophic scale
and shape parameters have been obtained and are displayed in Table \ref{tab:1}.
The result show that when a specific $\beta_{\scriptscriptstyle N}$ is given, increasing the $\alpha_{\scriptscriptstyle N}$ values results in higher qualitative characteristic values. On the other hand, with a specific $\alpha_{\scriptscriptstyle N}$, larger $\beta_{\scriptscriptstyle N}$ values lead to higher $E$ and $V$, while the values of $CV, Sk,$ and $Ku$ remain unchanged. This is because, as observed in Table \ref{SC}, these statistical characteristics are independent of $\beta_{\scriptscriptstyle N}$.

\begin{table}[!htbp]
    \centering
    \caption{The mean, variance, coefficient of variation, skewness and kurtosis for a selection of values of the neutrosophic parameters}
    \label{tab:1}
    \resizebox{\linewidth}{!}{
    \begin{tabular}{ccccccc}
    \hline 
    $\beta_{\nN}$ & $\alpha_{\nN}$ & E& V& CV & Sk& Ku\\
    \hline 
    \multirow{4}{*}{$[0.5, 1]$}
     & $[0.1, 0.35]$ & $[0.502, 1.061]$ & $[0.003, 0.141]$ & $[0.1, 0.354]$ & $[0.3, 1.038]$ & $[3.15, 4.775]$ \\
     & $[0.5, 0.75]$ & $[0.562, 1.281]$ & $[0.082, 0.958]$ & $[0.509, 0.764]$ & $[1.455, 2.056]$ & $[6.442, 9.713]$\\
     & $[1, 1.5]$ & $[0.75, 2.125]$ & $[0.562, 8.578]$ & $[1, 1.378]$ & $[2.519, 3.098]$ & $[12.852, 17.469]$\\
     & $[2, 3]$ & $[1.5, 5.5]$ & $[6, 110.25]$ & $[1.633, 1.909]$ & $[3.402, 3.673]$ & $[20.167, 22.724]$\\
    \hline 
     \multirow{4}{*}{$[1,2]$} & $[0.1, 0.35]$ & $[1.005, 2.122]$ & $[0.01, 0.565]$ & $[0.1, 0.354]$ & $[0.3, 1.038]$ & $[3.15, 4.775]$\\
     & $[0.5, 0.75]$ & $[1.125, 2.562]$ & $[0.328, 3.832]$ & $[0.509, 0.764]$ & $[1.455, 2.056]$ & $[6.442, 9.713]$ \\
     & $[1, 1.5]$ & $[1.5, 4.25]$ & $[2.25, 34.312]$ & $[1, 1.378]$ & $[2.519, 3.098]$ & $[12.852, 17.469]$\\
     & $[2, 3]$ & $[3, 11]$ & $[24, 441]$ & $[1.633, 1.909]$ & $[3.402, 3.673]$ & $[20.167, 22.724]$\\
    \hline 
    \multirow{4}{*}{$[2,3]$} & $[0.1, 0.35]$ & $[2.01, 3.184]$ & $[0.041, 1.271]$ & $[0.1, 0.354]$ & $[0.3, 1.038]$ & $[3.15, 4.775]$\\
    & $[0.5, 0.75]$ & $[2.25, 3.844]$ & $[1.312, 8.622]$ & $[0.509, 0.764]$ & $[1.455, 2.056]$ & $[6.442, 9.713]$\\
    & $[1, 1.5]$ & $[3, 6.375]$ & $[9, 77.203]$ & $[1, 1.378]$ & $[2.519, 3.098]$ & $[12.852, 17.469]$\\
	& $[2, 3]$ & $[6, 16.5]$ & $[96, 992.25]$ & $[1.633, 1.909]$ & $[3.402, 3.673]$ & $[20.167, 22.724]$\\
    \hline 
    \end{tabular}   
    }
\end{table}

\section{Neutrosophic parametric estimation}\label{sec:4}
In this section in order to investigate the effectiveness of parametric estimation, we first briefly review the maximum likelihood estimator (MLE) of BSD's parameters, as incorporated by Birnbaum and Saunders \cite{birnbaum1969estimation}. We then discuss the neutrosophic maximum likelihood estimator (NMLE) of the NBSD's parameters. 
The log-likelihood function is given by
\begin{align}\label{L}
     l(\alpha, \beta \mid D) \propto & -n \ln (\alpha)-n \ln (\beta) \nonumber\\ &+\sum_{i=1}^n \ln \left[\left(\frac{\beta}{t_i}\right)^{1 / 2}
     +\left(\frac{\beta}{t_i}\right)^{3 / 2}\right]-\frac{1}{2 \alpha^2} \sum_{i=1}^n\left(\frac{t_i}{\beta}+\frac{\beta}{t_i}-2\right),
\end{align}
where $D = \{t_1, \cdots, t_n\}$ is a random sample following $BS(\alpha, \beta)$.
To maximize \eqref{L}, by solving $\frac{\partial}{\partial\alpha} l(\alpha, \beta \mid D) = 0$, we obtain
\begin{equation}\label{MLalpha}
    \alpha^2=\left[\frac{s}{\beta}+\frac{\beta}{r}-2\right],
\end{equation}
where $s$ and $r$ are  the arithmetic and harmonic means of the sample as follows 
$$
s=\frac{1}{n} \sum_{i=1}^n t_i \quad \text { and } \quad r=\left[\frac{1}{n} \sum_{i=1}^n t_i^{-1}\right]^{-1}.
$$
Next, by differentiating \eqref{L} with respect to $\beta$ and equating it to zero, after substituting $\alpha^2$ using \eqref{MLalpha}, the non-linear equation is obtained as follows
\begin{equation}\label{DLbeta}
    \beta^2-\beta(2 r+K(\beta))+r(s+K(\beta))=0,
\end{equation}
where
$$
K(x)=\left[\frac{1}{n} \sum_{i=1}^n\left(x+t_i\right)^{-1}\right]^{-1} \quad \text { for } x \geq 0 .
$$
By solving the non-linear equation \eqref{DLbeta} through numerical iterative procedures, $\hat{\beta}$ is computed. The positive root of \eqref{DLbeta} is the MLE of $\beta$, say $\hat{\beta}$. Birnbaum and Saunders\cite{birnbaum1969estimation} also indicated $\hat{\beta}$ is the only positive root of \eqref{DLbeta} and moreover, $r<\hat{\beta}<s$. They proposed two different iterative numerical methods and showed that both methods converge to $\hat{\beta}$  using any optional initial guess value in $(r, s)$. After computing $\hat{\beta}$, the MLE of $\alpha$ is equal to

\begin{equation}\label{MLa}
    \hat{\alpha}=\left[\frac{s}{\hat{\beta}}+\frac{\hat{\beta}}     
     {r}-2\right]^{1 / 2}.
\end{equation}

Furthermore, they showed that the MLEs of $\alpha$ and $\beta$  are consistent estimators of the parameters.  Balakrishnan and Zhu \cite{balakrishnan2014existence} proved that if $n=1$, the MLEs do not exist, while for $n>1$, the MLEs are unique.

Accordingly, we consider the indeterminacy within the dataset, treating it as a continuous interval, with each observation denoted as $t_{\scriptscriptstyle Ni} = (t_{\scriptscriptstyle Li}, t_{\scriptscriptstyle Ui}), ~ \forall i=1, \cdots, n$. Hence, the likelihood function is given as
\begin{equation}\label{LN}
     l(\alpha_{\scriptscriptstyle N}, \beta_{\scriptscriptstyle N} \mid D_{\scriptscriptstyle N}) \propto \prod_{i=1}^{n} f(t_{{\scriptscriptstyle N}i}; \alpha_{\scriptscriptstyle N}, \beta_{\scriptscriptstyle N}), 
\end{equation}
where $D_{\scriptscriptstyle N} = \{t_{\scriptscriptstyle N1}, \cdots, t_{\scriptscriptstyle Nn}\}$ is a random sample following $NBS(\alpha_{\scriptscriptstyle N}, \beta_{\scriptscriptstyle N})$.
It is evident that in neutrosophic statistics, neutrosophic observations lead to neutrosophic parameters estimates. Thus, to derive the NMLEs, we need to determine the optimal neutrosophic intervals for parameters that maximize the likelihood function. This objective can be achieved by exploring the indeterminacy space within the data, denoted as $S_{\scriptscriptstyle N}=\prod_{i=1}^{n} t_{\scriptscriptstyle Ni}$. For all $\underline{t}=(t_1, \cdots, t_n) \in S_{\scriptscriptstyle N}$, MLEs of parameters can be computed with respect to (\ref{DLbeta}) and (\ref{MLa}). Therefore, it is necessary to calculate all MLEs within $S_{\scriptscriptstyle N}$ and identify the boundaries of the estimates. Let us denote the NMLE of parameters as ${\hat\alpha}_{\scriptscriptstyle N}=({\hat\alpha}_{\scriptscriptstyle NL}, {\hat\alpha}_{\scriptscriptstyle NU})$ and ${\hat\beta}_{\scriptscriptstyle N}=({\hat\beta}_{\scriptscriptstyle NL}, {\hat\beta}_{\scriptscriptstyle NU})$, where the components can be obtained as follows
\begin{eqnarray}
\hat \alpha_{\scriptscriptstyle NL} &= \underset{\underline{t} \in S_{\scriptscriptstyle N}}{\operatorname{min}} (\hat \alpha | \hat \alpha \in \underset{(\alpha, \beta)}{\operatorname{argmax}} (l(\alpha, \beta | \underline{t})), ~~~
\hat \alpha_{\scriptscriptstyle NU} = \underset{\underline{t} \in S_{\scriptscriptstyle N}}{\operatorname{max}} (\hat \alpha | \hat \alpha \in \underset{(\alpha, \beta)}{\operatorname{argmax}} (l(\alpha, \beta | \underline{t})),\label{NML1}\\
\hat \beta_{\scriptscriptstyle NL} &= \underset{\underline{t} \in S_{\scriptscriptstyle N}}{\operatorname{min}} (\hat \beta | \hat \beta \in \underset{(\alpha, \beta)}{\operatorname{argmax}} (l(\alpha, \beta | \underline{t})), ~~~
\hat \beta_{\scriptscriptstyle NU} = \underset{\underline{t} \in S_{\scriptscriptstyle N}}{\operatorname{max}} (\hat \beta | \hat \beta \in \underset{(\alpha, \beta)}{\operatorname{argmax}} (l(\alpha, \beta | \underline{t})).\label{NML2}
\end{eqnarray}
Therefore, the NMLEs of parameters are computed by determining the lower and upper bounds of the estimates as outlined in (\ref{NML1}) and (\ref{NML2}).

\section{Simulation study}\label{sec:5}

In this section, a simulation study is presented to check the performance of the NMLEs in NBSD.
As we know, 
in classical statistics, we deal with precise values, while in neutrosophic statistics, observations come with indeterminacy which can occur in any form depending on context. 
According to Smarandache \cite{Smarandache:2014}, indeterminacy and randomness are distinct concepts.
 Indeterminacy is due to the defects of the construction of
 the physical space (where an event can occur), and/or to
 the imperfect construction of the physical objects
 involved in the event, etc.
 Therefore, neutrosophic probability analyzes both the randomness and the indeterminacy related to
 the data. In this regard, Khan Sherwani et al. \cite{khan2021neutrosophic} considered a neutrosophic variable as
$$
T_{\scriptscriptstyle{N}} = T + I 
$$
where, both $T$ and $I$ are real numbers, which $T$ is the determined part, and $I$ shows the uncertain part of the neutrosophic variable. 
This form considered as an extension of classic random variable where $T$ is a real valued precise random variable stemming form the classical approach, and $I$ can be constant or a variable that follows a uniform distribution. 
Hence, to run the simulation, we generate $\mathrm{N}=$1000 random samples of sizes $n=$30, 50, 100, and 200 from NBSD. The desired neutrosophic datasets are built regarding the random values of  $I$ as $I \sim U(-\epsilon, \epsilon)$; $\epsilon =$ 0.005, 0.01 and 0.1.

Similar to finding NMLEs in (\ref{NML1}) and (\ref{NML2}), we compute neutrosophic average estimates (NAE), neutrosophic average biases (NAB) and neutrosophic measure square errors (NMSE) by defining their minimum and maximum values based on the dataset, and report them in Table \ref{tab:2}. 
As expected, results show that the NMSE reduces as the sample size increases. In addition, increasing the values of indeterminacy ($\epsilon$) leads to the greater uncertainty in the NAE intervals.
For example, when $\alpha = 0.5, \beta = 1.5,$ and $n = 200$, NAE values for $\alpha_{\scriptscriptstyle N}$ are:
\begin{align*}
\epsilon = 0.005 &: ~~~~~~~\underleftrightarrow{[0.4846, 0.486]}\\
\epsilon = 0.01 &: ~~~~~\underleftrightarrow{[0.4836 ~~,~~ 0.487]}\\
\epsilon = 0.1 &: ~~~\underleftrightarrow{[0.469       ~~~~~,~~~~~       0.5045]}
\end{align*}
so that as indeterminacy increases, the expanded intervals will be allocated to AE.

\afterpage{
\begin{landscape}
\renewcommand{\arraystretch}{1.1}
\begin{table}[htbp]
\caption{NMLEs simulation}\label{tab:2}
\resizebox{\linewidth}{!}{
\begin{tabular}{ccccccccccc}
\hline 
  &&& \multicolumn{2}{c}{ NAE } & & \multicolumn{2}{c}{ NAB } && \multicolumn{2}{c}{ NMSE } \\
\cline { 4-5 } \cline { 7-8} \cline { 10-11 } 
&$\epsilon$ &  $n$ & $\hat{\alpha}_{N}$ & $\hat{\beta}_{N}$ && $\hat{\alpha}_{N}$ & $\hat{\beta}_{N}$ && $\hat{\alpha}_{N}$ & $\hat{\beta}_{N}$ \\
\hline
$\alpha = 0.25$, $\beta = 1$ &&&&&&&&&&\\
\hline
&0.005 &30 & [0.2419, 0.2437] & [0.9980, 1.0002] && [-0.0081, -0.0063] & [-0.002, 2e-04] && [0.0315, 0.0319] & [0.0448, 0.0449]\\
&&50 & [0.244, 0.2458] & [0.9992, 1.0013] && [-0.006, -0.0042] & [-8e-04, 0.0013] && [0.0236, 0.024] & [0.0354, 0.0354]\\
&&100 & [0.2473, 0.2491] & [0.9997, 1.0018] && [-0.0027, -9e-04] & [-3e-04, 0.0018] && [0.0175, 0.0176] & [0.0248, 0.0248]\\
&&200 & [0.2484, 0.2502] & [0.9985, 1.0007] && [-0.0016, 2e-04] & [-0.0015, 7e-04] && [0.0128, 0.0128] & [0.0170, 0.0170]\\
\hline  
&0.01 &30 & [0.2407, 0.2449] & [0.9966, 1.0015] && [-0.0093, -0.0051] & [-0.0034, 0.0015] && [0.0313, 0.0322] & [0.0449, 0.0449]\\
&&50 & [0.2428, 0.2470] & [0.9978, 1.0027] && [-0.0072, -0.003] & [-0.0022, 0.0027] && [0.0235, 0.0243] & [0.0354, 0.0355]\\
&&100 & [0.2461, 0.2503] & [0.9982, 1.0032] && [-0.0039, 3e-04] & [-0.0018, 0.0032] && [0.0175, 0.0178] & [0.0248, 0.025]\\
&&200 & [0.2472, 0.2514] & [0.9971, 1.0021] && [-0.0028, 0.0014] & [-0.0029, 0.0021] && [0.0128, 0.013] & [0.0171, 0.0172]\\
\hline
&0.1 &30 & [0.2241, 0.2677] & [0.9728, 1.0241] && [-0.0259, 0.0177] & [-0.0272, 0.0241] && [0.0364, 0.0402] & [0.0514, 0.0527]\\
&&50 & [0.2265, 0.2701] & [0.9736, 1.0252] && [-0.0235, 0.0201] & [-0.0264, 0.0252] && [0.0313, 0.0329] & [0.0434, 0.0441]\\
&&100 & [0.2297, 0.2735] & [0.9739, 1.0258] && [-0.0203, 0.0235] & [-0.0261, 0.0258] && [0.0268, 0.0297] & [0.0359, 0.0361]\\
&&200 & [0.2308, 0.2744] & [0.973, 1.0248] && [-0.0192, 0.0244] & [-0.027, 0.0248] && [0.023, 0.0277] & [0.0302, 0.032]\\
\hline

$\alpha = 0.5$,  $\beta = 1.5$ &&&&&&&&&\\ 
\hline 
& 0.005 & 30&  [0.4846,  0.4860] & [1.4990, 1.5014] &&[-0.0154,  -0.0140] &[-0.0010, 0.0014] && [0.0633, 0.0635] & [0.1316,  0.1316]\\
&& 50 &  [0.4887,  0.4902] & [1.5013,  1.5037]&&[-0.0113, -0.0098] & [0.0013,  0.0037]&& [0.0476,  0.0478] & [0.1043,  0.1044]\\
&& 100 & [0.4955,  0.4970] & [1.5016,  1.5040] && [-0.0045,  -0.0030] & [0.0016,  0.0040] && [0.0350,  0.0351] & [0.0728,  0.0729] \\
&& 200  & [0.4978,  0.4993] & [1.4981,  1.5005] && [-0.0022, -7e-04] &[-0.0019,  5e-04] && [0.0255,  0.0256] & [0.0498,  0.0498]\\
\hline
& 0.01 & 30& [0.4836,  0.4870] & [1.4974, 1.5029] && [-0.0164, -0.0130]&[-0.0026, 0.0029] &&[0.0631,  0.0637] & [0.1316,  0.1317] \\
&& 50 & [0.4878,  0.4912] & [1.4997,  1.5052] && [-0.0122,  -0.0088] & [-3e-04,  0.0052] && [0.0474,  0.0480] & [0.1043,  0.1044] \\
&& 100 & [0.4945,  0.4980] & [1.5000,  1.5056] && [-0.0055,  -0.0020] & [0.0000,  0.0056] && [0.0350,  0.0352] & [0.0728,  0.0730] \\
&& 200  & [0.4968,  0.5002] & [1.4965,  1.5020] && [-0.0032,  2e-04] & [-0.0035,  0.0020] && [0.0256,  0.0257] & [0.0499,  0.0499]\\
\hline
&0.1 &30 & [0.4690, 0.5045] & [1.4708, 1.5279] && [-0.0310, 0.0045] & [-0.0292, 0.0279] && [0.0635, 0.0683] & [0.1350, 0.1352]\\
&&50 & [0.4733, 0.5091] & [1.4725, 1.5301] && [-0.0267, 0.0091] & [-0.0275, 0.0301] && [0.0486, 0.0530] & [0.1083, 0.1076]\\
&&100 & [0.4799, 0.5160] & [1.4726, 1.5307] && [-0.0201, 0.0160] & [-0.0274, 0.0307] && [0.0394, 0.0400] & [0.0780, 0.0790]\\
&&200 & [0.4822, 0.5181] & [1.4693, 1.5273] && [-0.0178, 0.0181] & [-0.0307, 0.0273] && [0.0308, 0.0320] & [0.0571, 0.0588]\\
\hline
\end{tabular}
}
\end{table}
\end{landscape}
}

\section{Applications}\label{sec:6}
\subsection{Pseudo neutrosophic data}

In this section, we will explore the application of the NBS model to real datasets.
We utilize failure life data, where the dataset aligns with the characteristics of the BS distribution explored by   Birnbaum and Saunders \cite{birnbaum1969new}. The dataset, comprising 101 observations, represents the fatigue life of 6061-T6 aluminium coupons. These coupons were cut in parallel to the rolling direction and exposed to oscillations at a frequency of 18 cycles per second (cps). It's noteworthy that all entries in the dataset share a consistent maximum stress per cycle of 31,000 psi. The dataset is provided in Table \ref{Flac}.

\begin{table}[!hbp]
\caption{Fatigue life of 6061-T6 aluminium coupons}
\label{Flac}
\centering
\resizebox{\linewidth}{!}{
\begin{tabular}{c}
\hline
70 90 96 97 99 100 103 104 104 105 107 108 108 108 109 109 112 112 113 114 114 114\\
116 119 120 120 120 121 121 123 124 124 124 124 124 128 128 129 129 130 130 130 131 131\\
131 131 131 132 132 132 133 134 134 134 134 134 136 136 137 138 138 138 139 139 141 141\\
142 142 142 142 142 142 144 144 145 146 148 148 149 151 151 152 155 156 157 157 157 157\\
158 159 162 163 163 164 166 166 168 170 174 196 212 \\
\hline
\end{tabular}
}
\end{table}

From Balakrishnan and Kundu \cite{balakrishnan2019birnbaum}, we have $s=(1 / n) \sum_{i=1}^n t_i=133.73267$, and $r=n / \sum_{i=1}^n t_i^{-1}$ $=$ 129.93321. Point estimates of $\alpha$ and $\beta$ are $\hat\alpha=0.170385$ and $\hat\beta=131.818792$.

In our paper to effectively demonstrate the practical application of a neutrosophic context, we employed the specialized method described in Section \ref{sec:5} of the paper. This method allows for incorporation of indeterminacy, a critical consideration in situations where uncertainty or vagueness is intrinsic.

To explore the impact of varying levels of indeterminacy (I) on our parameter estimates, we introduced different values of $\epsilon$, as $\epsilon$ = 0, 0.001, 0.01, and 0.1. Each $\epsilon$ value represents a distinct degree of indeterminacy, allowing us to observe the model's responsiveness to different levels of uncertainty.
The resulting parameter estimates, which include $\alpha$ and $\beta$, are detailed in Table \ref{tab2}. This tabular representation provides a clear overview of how the model's parameters change under different degrees of indeterminacy.
Moreover, to completely assess the model's performance, we have included the corresponding values for Akaike’s Information Criteria (AIC), Bayesian Information Criteria (BIC), and goodness-of-fit test in Table \ref{tab2}. 
 Consider that within the neutrosophic concept, all relevant values are calculated by identifying the minimum and maximum values in the dataset.

It is well-known that when the parameters of a distribution are not known, one should apply the modified goodness-of-fit tests such as Anderson-Darling (AD), Cramér-von Mises (CM), and Kolmogorov-Smirnov (KS) introduced by  D'Agostino and Stephens \cite{1986goodness}. Here, we are opting for the modified KS test where the statistic for a given dataset $t_1, \cdots, t_n$ following distribution $F$ is

\begin{equation}\label{KS}
KS^*=\left[\sqrt{n}-0.01+\frac{0.85}{\sqrt{n}}\right] \max \left\{D^{+}, D^{-}\right\},
\end{equation}
\begin{equation*}
D^{+}= \max _{j=1, \ldots, n}\{j / n-u_j\}, ~~ \text{and} ~~
D^{-}= \max _{j=1, \ldots, n}\{u_j-[j-1] / n\}.
\end{equation*}

To compute the test results, we're employing an algorithm proposed by Chen and Balakrishnan \cite{chen1995general} for skewed distributions. The procedure is summarized in algorithm \ref{KS;al}. 

\begin{algorithm}
    \caption{Modified KS test}\label{KS;al}
    \begin{algorithmic}
                \STATE 1. Compute MLEs of $\alpha_{\scriptscriptstyle N}$ and $\beta_{\scriptscriptstyle N}$;
                \STATE 2. Compute $v_i=F(t_i ; \alpha_{\scriptscriptstyle N}, \beta_{\scriptscriptstyle N})$, where the $t_i$s are in ascending order;
                \STATE 3. Compute $y_i=\Phi^{-1}\left(v_i\right)$, where $\Phi$ is the standard normal CDF and $\Phi^{-1}$ is its inverse;
                \STATE 4. Compute $u_i=\Phi\left\{\left(y_i-\bar{y}\right) / s_y\right\}$, where $\bar{y}=$ $n^{-1} \sum_{i=1}^n y_i$ and $s_y^2=(n-1)^{-1} \sum_{i=1}^n\left(y_i-\bar{y}\right)^2$;
                \STATE 5. Calculate $KS^*$ according to \eqref{KS}, 
                \STATE 6. Compare the $\mathrm{KS}^*$ statistic evaluated in step 5 with the corresponding quantiles of their distributions, and determine the respective $p$-values.
				\STATE 7. Decide whether the dataset follows the NBSD or not for a significance level specified.
    \end{algorithmic}
\end{algorithm}

Values in Table \ref{tab2} help us measure and compare how well a model fits and performs across different uncertainty levels.
According to the results, it's clear that when uncertainty decreases, the model tends to exhibit characteristics closer to classical models associated with certainty.

\subsection{Environmental dataset}
Neutrosophic logic can be applied to model and analyze environmental datasets where uncertainties and incomplete information are really common. This includes climate modeling, ecological studies, and environmental risk assessment. Environmental datasets often exhibit uncertainties due to various factors such as measurement errors, variability, and incomplete information. 
Neutrosophic statistics enables us to effectively model and depict uncertainty, facilitating a more nuanced comprehension of the data.
For example, in modeling the indeterminacy in climate model parameters and dealing with incomplete observational data, or in hydrological models which often involve uncertain parameters and incomplete information about rainfall, runoff, and other hydrological variables, even in environmental emergencies or disasters, where information may be rapidly changing and incomplete.

\begin{table}[t]
    \renewcommand{\arraystretch}{1.2}
    \caption{Parameters estimates and goodness-of-fit statistics for aluminium coupons dataset. }\label{tab2}  
    \resizebox{\linewidth}{!}{
 	\begin{tabular}{ccccccc}
 			\hline 
 			$\epsilon$ & Parameter & Estimate &  Log-likelihood & AIC & BIC & $\mathrm{KS}^*$ \\
 			\hline
             \multirow{2}{*}{0} & $\hat\alpha_{\scriptscriptstyle N}$  & 0.170385   
             & \multirow{2}{*}{-457.270528} & \multirow{2}{*}{918.541056} & \multirow{2}{*}{933.001538} & \multirow{2}{*}{0.8577896} \\
             & $\hat\beta_{\scriptscriptstyle N}$ & 131.818792
             & & & & \\
             \hline
             \multirow{2}{*}{0.001} & $\hat\alpha_{\scriptscriptstyle N}$  & [0.170379, 0.170391]   
             & \multirow{2}{*}{[-457.274, -457.266]} & \multirow{2}{*}{[918.533, 918.548]} & \multirow{2}{*}{[932.994, 933.008]} & \multirow{2}{*}{[0.857,  0.857]} \\
             & $\hat\beta_{\scriptscriptstyle N}$ & [131.817, 131.819]
             & & & & \\
             \hline
             \multirow{2}{*}{0.01} & $\hat\alpha_{\scriptscriptstyle N}$  & [0.170323, 0.170446]   
             & \multirow{2}{*}{[-457.307, -457.233]} & \multirow{2}{*}{[918.467, 918.614]} & \multirow{2}{*}{[932.928, 933.074]} & \multirow{2}{*}{[0.856,  0.858]} \\
             & $\hat\beta_{\scriptscriptstyle N}$ & [131.808, 131.828]
             & & & & \\
             \hline
             \multirow{2}{*}{0.1} & $\hat\alpha_{\scriptscriptstyle N}$  & [0.169771, 0.171]   
             & \multirow{2}{*}{[-457.636, -456.903]} & \multirow{2}{*}{[917.807, 919.273]} & \multirow{2}{*}{[932.268, 933.733]} & \multirow{2}{*}{[0.848,  0.866]} \\
             & $\hat\beta_{\scriptscriptstyle N}$ & [131.717, 131.920]
             & & & & \\
             \hline
             \multirow{2}{*}{1} & $\hat\alpha_{\scriptscriptstyle N}$  & [0.164316,  0.176615]   
             & \multirow{2}{*}{[-460.916,  -453.589]} & \multirow{2}{*}{[911.178,  925.832]} & \multirow{2}{*}{[925.638,  940.293]} & \multirow{2}{*}{[0.754,  0.940]} \\
             & $\hat\beta_{\scriptscriptstyle N}$ & [130.802,  132.834]
             & & & & \\
 			\hline
        \end{tabular}
 	}
\end{table}

To demonstrate the computational methodology of the proposed NBS model, we turn to a tangible environmental dataset used by Khan et al. \cite{khan2021statistical} on yearly Nitrogen oxides emissions in Denmark. This dataset covers 1990 to 2018 and has been thoroughly calculated by the United Nations Statistics Divisions (UNSD), accessible on their website. The dataset is presented in Table \ref{Noe}.
\begin{table}[!tbp]
\centering
\caption{Nitrogen oxides emissions dataset in Denmark 1990-2018}
\label{Noe}
\resizebox{\linewidth}{!}{
\begin{tabular}{c}
\hline
$[304.12, 307.82], 355.34, 310.93, 309.47, [ 309.12, 312.10], 292.80, 327.49, 280.33, 259.99,$\\
$ [238.19, 242.45], 229.98, 226.77, 223.57, 233.12,
216.37, 208.16, [206.30, 209.14], 193.44, 177.31, $\\
$157.88, 153.18,
143.93, 132.78, 127.87, 118.28, 116.75, 116.96, 114.21,
[106.86, 110.62]$  \\
\hline
\end{tabular}
}
\end{table}
The determination of Nitrogen oxides emissions per capita involves an established international methodology, incorporating country-specific details related to industrial, energy, waste management, and agricultural production.

The dataset reveals a noteworthy characteristic: Nitrogen oxides emissions are reported not as precise values but as intervals, containing [304.12, 307.82], [309.12, 312.10], [238.19, 242.45], [206.30, 209.14], and [106.86, 110.62]. 
The imprecise emissions measurement present a challenge for classical models like the BS model, making them ineffective due to ambiguity and uncertainties in the data. In contrast, the proposed neutrosophic model easily handles a set of measurements with uncertainties, offering a more nuanced and robust analysis. 

We first calculate $\mathrm{KS}^*$ value according to Algorithm \ref{KS;al} and the associated p-value through  Monte Carlo (MC) simulation method. Subsequently, we conduct a comparative analysis.
Given the inherent ambiguity and uncertainty present in neutrosophic datasets, relying solely on classical quantitative metrics is not feasible. Classical statistical tests are designed under the assumption of precise and well-defined data, which contrasts with the fundamental nature of neutrosophic data. Attempting to apply classical analysis methods to neutrosophic datasets would overlook the essential characteristics of ambiguity and indeterminacy. However, this does not hinder our ability to compare the performance of different neutrosophic models. Here, we focus on comparing the NBSD with two other neutrosophic models: the neutrosophic gamma (NG) and neutrosophic log-normal (NLN) distributions, introduced by Khan et al. \cite{khan2021statisticalg, khan2021statistical}. The results are summarized in Table \ref{Nit}.
Since all models have p-values higher than the significance level (e.g., $\alpha = 0.05$), indicating that they all fit the data adequately according to the goodness-of-fit test, we should prefer the model with the lowest AIC and BIC values.
It is obvious that the NBSD is preferred because it has the lowest AIC and BIC values among the three models, indicating the best balance between goodness-of-fit and model complexity. 

\begin{table}[!ht]
    \centering
    \caption{ Parameter estimates and goodness-of-fit statistics for models of the environmental dataset. }\label{Nit}
 	\resizebox{\linewidth}{!}{
 	\begin{tabular}{ccccccc}
 			\hline
			Model & MLEs &  Log-likelihood & AIC & BIC &  $\mathrm{KS}^*$ & p-value\\
			\hline

			\multirow{2}{*}{NBSD} & 
			$\hat\alpha_{\scriptscriptstyle N}=$ [0.3702,0.3736]&
			\multirow{2}{*}{  [-165.8135, -165.544] }&  \multirow{2}{*}{[335.0881, 335.6269]} & \multirow{2}{*}{[344.5573, 345.0961]}&\multirow{2}{*}{ [0.659, 0.6939]}& \multirow{2}{*}{[0.2786, 0.3628]}\\
			&$\hat\beta_{\scriptscriptstyle N}=$   [199.6423, 200.2666] &&&&&
			 \\	
            \hline

            \multirow{2}{*}{NLND} & 
			$\hat\mu_{\scriptscriptstyle N}=$[5.2985, 5.3015]	&
			\multirow{2}{*}{ [-166.0176, -165.747] }&  \multirow{2}{*}{[335.4941, 336.0351]} & \multirow{2}{*}{[344.9632, 345.5043] }&\multirow{2}{*}{ [0.6581, 0.6934]}& \multirow{2}{*}{ [0.2798, 0.3648]}\\
			&$\hat\sigma_{\scriptscriptstyle N}=$[0.3667, 0.3699] &&&&& \\	
            \hline

            \multirow{2}{*}{NGD} & 
			$\hat\alpha_{\scriptscriptstyle N}=$[0.0784,0.0784]	&
			\multirow{2}{*}{ [-234.6191, -234.5202]  }&  \multirow{2}{*}{[473.0404, 473.2382]} & \multirow{2}{*}{[482.5096, 482.7074]  }&\multirow{2}{*}{ [0.5734, 0.5900]}& \multirow{2}{*}{ [0.5377, 0.5811]}\\
			&$\hat\sigma_{\scriptscriptstyle N}=$[2722.3376, 2731.3692] &&&&& \\	
            \hline
        \end{tabular}
        }
\end{table}

Consequently, utilizing neutrosophic statistical models for assessment, monitoring, and forecasting of environmental data can provide more accurate and timely predictions, which are crucial for the mitigation of environmental-related impacts.

\section{Conclusion}\label{sec:7}

This research presents the neutrosophic Birnbaum-Saunders distribution (NBSD), a probabilistic model tailored for neutrosophic analysis applications. While the classical Birnbaum-Saunders distribution remains suitable for well-defined datasets, the neutrosophic extension offers a robust alternative for handling the complexities of more ambiguous datasets. Utilizing R software, we conduct both simulated studies and real data analysis, demonstrating the practicality and adaptability of our model in addressing real-world challenges. The model has been implemented in real-world industrial and environmental applications to verify its practical significance. The results demonstrate the effectiveness of the NBSD model when compared to the neutrosophic log-normal and gamma models.

\section*{Declaration of competing interest}
The authors declare that they have no known competing financial interests or personal
relationships that could have appeared to influence the work reported in this paper.

\section*{Acknowledgments}
Mohammad Arashi’s work is based on the research supported in part by the Iran
National Science Foundation (INSF) grant No. 4015320.


\begin{thebibliography}{99}

\bibitem{Smarandache:2014}
F. Smarandache, Introduction to Neutrosophic Statistics, Sitech \& Education
Publishing., 2014.

\bibitem{Smarandache:1998}
F. Smarandache, Neutrosophy: Neutrosophic Probability, Set, and Logic: Analytic
Synthesis \& Synthetic Analysis, American Research Press, Rehoboth, NM,
1998.

\bibitem{Smarandache:2005}
F. Smarandache, A generalization of the intuitionistic fuzzy set, International
Journal of Pure and Applied Mathematics 24 (3) (2005) 287-297.

\bibitem{Smarandache:2007} 
F. Smarandache, A unifying field in logics: neutrosophic logic. First International
Conference on Neutrosophy, Neutrosophic Logic, Set, Probability, and Statistics,
6th Edition, InfoLearnQuest, USA, 2007.

\bibitem{Smarandache:2010}
F. Smarandache, Neutrosophic logic-a generalization of the intuitionistic fuzzy
logic, Multispace \& multistructure. Neutrosophic Transdisciplinarity (100 collected
papers of science) 4 (2010) 396.

\bibitem{Smarandache:2016} 
F. Smarandache, Neutrosophic Overset, Neutrosophic Underset, and Neutrosophic
Offset. Similarly for Neutrosophic Over-/Under-/Off-Logic, Probability,
and Statistics, Brussels, Belgium, EU: Pons Editions., 2016.

\bibitem{Smarandache:Pramanik:2016}
F. Smarandache, S. Pramanik, New Trends in Neutrosophic Theory and Applications,
Vol. 1, Brussels, Belgium, EU: Pons Editions., 2016.

\bibitem{Patro:Smarandache:2016}
S. K. Patro, F. Smarandache, The neutrosophic statistical distribution, more
problems, more solutions, Neutrosophic Sets and Systems 12 (2016) 73-79.

\bibitem{Ali:2018}
M. Ali, L. Q. Dat, L. H. Son, F. Smarandache, Interval complex neutrosophic
set: Formulation and applications in decision-making, International Journal of
Fuzzy Systems 20 (2018) 986-999.

\bibitem{Salama:Alblowi:2012}
A. Salama, S. Alblowi, Generalized neutrosophic set and generalized neutrosophic
topological spaces, Computer Science and Engineering 2 (2012) 129-132.

\bibitem{Alhabib:2018}
R. Alhabib, M. M. Ranna, H. Farah, A. A. Salama, Some neutrosophic probability
distributions, Neutrosophic Sets and Systems 22 (2018) 30-38.

\bibitem{Alhasan:Smarandache:2019}
K. F. H. Alhasan, F. Smarandache, Neutrosophic weibull distribution and neutrosophic
family weibull distribution, Neutrosophic Sets and Systems 28 (2019)
191-199.

\bibitem{Khan:2021}
Z. Khan, M. Gulistan, N. Kausar, C. Park, Neutrosophic rayleigh model with
some basic characteristics and engineering applications, IEEE Access 9 (2021)
71277-71283.

\bibitem{khan2021statistical}
Z. Khan, A. Amin, S. A. Khan, M. Gulistan, Statistical development of the neutrosophic
lognormal model with application to environmental data, Neutrosophic
Sets and Systems 47 (1) (2021) 1.

\bibitem{Norouzirad:2023}
M. Norouzirad, G. S. Rao, D. Mazarei, Neutrosophic generalized rayleigh distribution
with application, Neutrosophic Sets and Systems 58 (1) (2023) 15.

\bibitem{Duan:2021}
W.-Q. Duan, Z. Khan, M. Gulistan, A. Khurshid, Neutrosophic exponential
distribution: modeling and applications for complex data analysis, Complexity
2021 (2021) 1-8.

\bibitem{rao2023neutrosophic}
G. S. Rao, M. Norouzirad, D. Mazarei, Neutrosophic generalized exponential
distribution with application, Neutrosophic Sets and Systems 55 (1) (2023) 28.

\bibitem{khan2021neutrosophic}
R. A. K. Sherwani, M. Naeem, M. Aslam, M. Raza, M. Abid, S. Abbas, Neutrosophic
beta distribution with properties and applications, Neutrosophic Sets
and Systems 41 (2021) 209-214.

\bibitem{Shah:2022}
F. Shah, M. Aslam, Z. Khan, M. Almazah, F. S. Alduais, et al., On neutrosophic
extension of the maxwell model: properties and applications, Journal of Function
Spaces 2022, article ID 4536260 (2022).

\bibitem{Ahsan-ul-Haq:2022}
M. Ahsan-ul Haq, Neutrosophic kumaraswamy distribution with engineering application,
Neutrosophic Sets and Systems 49 (2022) 269-276.

\bibitem{Eassa:2023}
N. I. Eassa, H. M. Zaher, N. A. A. El-Magd, Neutrosophic generalized pareto
distribution, Mathematics and Statistics 11 (2023) 827-833.

\bibitem{birnbaum1969new}
Z. W. Birnbaum, S. C. Saunders, A new family of life distributions, Journal of
Applied Probability 6 (2) (1969) 319-327.

\bibitem{balakrishnan2019birnbaum}
N. Balakrishnan, D. Kundu, Birnbaum-saunders distribution: A review of models,
analysis, and applications, Applied Stochastic Models in Business and Industry
35 (1) (2019) 4-49.

\bibitem{birnbaum1969estimation}
Z. W. Birnbaum, S. C. Saunders, Estimation for a family of life distributions with
applications to fatigue, Journal of Applied Probability 6 (2) (1969) 328-347.

\bibitem{balakrishnan2014existence}
N. Balakrishnan, X. Zhu, On the existence and uniqueness of the maximum
likelihood estimates of the parameters of birnbaum-saunders distribution based
on type-i, type-ii and hybrid censored samples, Statistics 48 (5) (2014) 1013-
1032.

\bibitem{Sherwani:2021}
R. A. Khan Sherwani, M. Naeem, M. Aslam, M. A. Raza, S. Abbas, et al.,
Neutrosophic beta distribution with properties and applications, Neutrosophic
Sets and Systems 41 (1) (2021) 12.

\bibitem{1986goodness}
C. D'Agostino, M. A. Stephens, Goodness of fit techniques, Marcel Dekker, Inc,
New York., 1986.

\bibitem{chen1995general}
G. Chen, N. Balakrishnan, A general purpose approximate goodness-of-fit test,
Journal of Quality Technology 27 (2) (1995) 154-161.

\bibitem{khan2021statisticalg}
Z. Khan, A. Al-Bossly, M. M. Almazah, F. S. Alduais, On statistical development
of neutrosophic gamma distribution with applications to complex data analysis,
Complexity 2021 (1) (2021) 3701236.

\end{thebibliography}

\end{document}